\documentclass[a4paper,fleqn,usenatbib]{mnras}

\usepackage{newtxtext,newtxmath}

\usepackage[T1]{fontenc}
\usepackage{ae,aecompl}

\usepackage{graphicx}	
\usepackage{amsmath}	
\usepackage{amssymb}	
\usepackage{xcolor}
\usepackage{subfig}
\usepackage{booktabs}

\title[]{First on-sky demonstration of an integrated-photonic nulling-interferometer: The GLINT instrument}

\author[Barnaby R. M. Norris, et al.]{
Barnaby R. M. Norris,$^{1,2,3}$\thanks{E-mail: barnaby.norris@sydney.edu.au}
Nick Cvetojevic,$^{4}$
Tiphaine Lagadec,$^{1,2,3}$
Nemanja Jovanovic,$^{5}$
\newauthor
Simon Gross,$^{6}$
Alexander Arriola,$^{6}$
Thomas Gretzinger,$^{6}$
Marc-Antoine Martinod,$^{1,2,3}$
\newauthor
Olivier Guyon,$^{7,8,9}$
Julien Lozi,$^{9}$
Michael J.Withford,$^{6}$
Jon S. Lawrence,$^{10}$
Peter Tuthill$^{1,2,3}$
\\
$^{1}$Sydney Institute for Astronomy, School of Physics, Physics Road, University of Sydney, NSW 2006, Australia\\
$^{2}$Sydney Astrophotonic Instrumentation Laboratories, Physics Road, University of Sydney, NSW 2006, Australia\\
$^{3}$AAO-USyd, School of Physics, University of Sydney 2006\\
$^{4}$Laboratoire Lagrange, Observatoire de la C\^{o}te d'Azur, Université C\^{o}te d'Azur, 06304 Nice, France\\
$^{5}$California Institute of Technology, 1200 E. California Blvd, Pasadena, 91125, CA, USA\\
$^{6}$MQ Photonics Research Centre, Department of Physics and Astronomy, Macquarie University, NSW,
2109, Australia\\
$^{7}$Astrobiology Center, National Institutes of Natural Sciences, 2-21-1 Osawa, Mitaka, Tokyo, JAPAN\\
$^{8}$Steward Observatory, University of Arizona, Tucson, AZ 85721, USA\\
$^{9}$National Astronomical Observatory of Japan, Subaru Telescope, National Institutes of Natural Sciences, Hilo, HI 96720, USA\\
$^{10}$Australian Astronomical Observatory, Faculty of Science and Engineering, Macquarie University, NSW 2109, Australia\\
}

\date{Accepted XXX. Received YYY; in original form ZZZ}

\pubyear{2019}

\begin{document}
\label{firstpage}
\pagerange{\pageref{firstpage}--\pageref{lastpage}}
\maketitle

\begin{abstract}
The characterisation of exoplanets is critical to understanding planet diversity and formation, their atmospheric composition and the potential for life. This endeavour is greatly enhanced when light from the planet can be spatially separated from that of the host star. One potential method is nulling interferometry, where the contaminating starlight is removed via destructive interference. The GLINT instrument is a photonic nulling interferometer with novel capabilities that has now been demonstrated in on-sky testing. The instrument fragments the telescope pupil into sub-apertures that are injected into waveguides within a single-mode photonic chip. Here, all requisite beam splitting, routing and recombination is performed using integrated photonic components. We describe the design, construction and laboratory testing of our GLINT pathfinder instrument. We then demonstrate the efficacy of this method on sky at the Subaru Telescope, achieving a null-depth precision on sky of $\sim10^{-4}$ and successfully determining the angular diameter of stars (via their null-depth measurements) to milli-arcsecond accuracy. A statistical method for analysing such data is described, along with an outline of the next steps required to deploy this technique for cutting-edge science. 
\end{abstract}




\section{Introduction}
With the detection of over 4000 exoplanets confirmed so far \citep{Schneider2011}, of increasing importance is the detailed characterisation of these planets. While indirect planet detection methods (such as transit and radial-velocity observations) have revolutionised our understanding of the ubiquity and diversity of exoplanets, the promise of \emph{directly imaging} exoplanets in the habitable-zone has been largely out of reach due to insurmountable observational challenges. However direct imaging is extremely attractive, as it produces measurements inaccessible to indirect methods, including direct measurement of orbital parameters, characterisation of surface features, weather, atmospheric composition and even the detection of signs of life \citep{Fujii2010, Kawahara2012, Snellen2014, Seager2016}. 

To obtain high signal to noise measurements of these properties, it is critical that the overwhelming starlight from the host star -- spatially separated from the planet by just tens of milli-arcseconds -- be suppressed, both to remove its photon noise and the (variable) effects of the stellar spectrum \citep{Guyon2012}.  
Required near-IR contrast ratios range from $\sim10^{-4}$ (for large self-luminous planets \citep{Marois2008}) to $\sim10^{-8}$ (for Earth-like planets reflecting their host-star's light \citep{Guyon2012, Schworer2015}). A young planet in a nearby star-forming region (at a distance of 100 parsecs) in the habitable zone of a sun-like star (1 AU orbit) will be at an angular separation of just 10 mas. Such planets are potentially within the resolving power of modern 8-metre telescopes, although the raw resolution (without adaptive-optics) of terrestrial telescopes is 100 times worse due to atmospheric seeing. To make matters worse, in the case of reflected-light planets there is a strong relationship between separation and contrast ratio due to the $1/r^2$ falloff in incident starlight, such that planets at their most favourable contrasts are at the smallest separations.

Conventional approaches to addressing this imaging challenge rely on the combination of wavefront correction via extreme adaptive optics and suppressing the starlight via a coronagraph (e.g. \cite{Jovanovic2015, Macintosh2014, Beuzit2008}). The performance of such systems is partly defined by the achievable inner-working-angle (IWA), the closest spatial separation from the central star that can be observed. Achieving IWAs better than 100 mas is as-yet extremely difficult; only modest suppression of starlight (and hence achievable contrast) is currently possible at the smallest IWAs, with performance rapidly increasing as the separation from the star increases. 

An alternative approach is the use of \emph{nulling interferometry}, first proposed by \cite{Bracewell1978}. As with conventional optical interferometers, light from separate regions of a telescope pupil (or separate telescopes) is brought together, and the resulting interference patterns analysed to deduce spatial information. However nulling interferometers also manipulate the phases of the individual beams such that the light interferes \emph{destructively} on-axis. The starlight is effectively `nulled' out (being redirected to regions of constructive interference elsewhere), and the faint, slightly off-axis planet-light remains. Nulling interferometry offers a key advantage over coronagraphic methods especially at very small star-planet separations. Unlike a coronagraph, nulling (and interferometry in general) has no fixed inner-working-angle. Rather, companions at separations at and beyond the formal diffraction limit can be observed, with the penalty of decreasing contrast sensitivity as the apparent star-planet separation becomes much smaller than the diffraction limit. 

Since the idea was originally proposed a wide range of implementations have been proposed (e.g. \cite{Serabyn2000}), including multiple combinations of baselines to allow high-resolution imaging \citep{Angel1997}, multi-element space-based instruments \citep{Leger1996}, and detection of exo-zodiacal disks \citep{Absil2006}. Several nulling interferometers have been built using conventional bulk-optics technologies, such as the Keck Interferometer Nuller \citep{Colavita2009} and the Large Binocular Telescope Interferometer \citep{Defrere2016}, but face a limitation in the maximum achievable starlight suppression. This arises from the fact that ideal cancellation could only be achieved if the wavefronts of the two beams were perfectly flat, and so the phase-difference at all points would be constant (and accordingly delayed by $\pi$ radians). However the actual wavefronts are anything but flat, due largely to atmospheric seeing as well as low order optical aberrations and optical surface roughness, greatly impacting the achievable performance \citep{Mennesson2002}. For the instruments mentioned above, operation at long wavelengths (8 -- 13$\mu$m) mitigated the severity of this effect, which becomes much more challenging at near-IR and visible wavelengths.

One solution is to implement spatial filtering via a single-mode fibre \citep{Foresto1990, Mennesson2002}. In this case, starlight from each telescope (or sub-aperture) is injected into a single-mode fibre or waveguide, which has the property that only the amplitude, global phase and polarisation of the light is transmitted. The resulting pure Gaussian beams are said to be ``filtered" so that all spatial substructure is lost (albeit at the expense of injection efficiency). In this case the resulting beams could form a perfectly deep null. In practice, the null depth would be limited by the ability to keep the phase delay between the two filtered beams constant despite seeing, as well as bandwidth and polarisation effects. The use of single-mode fibres in a nuller was demonstrated by the Palomar Fiber Nuller \citep{Mennesson2011, Kuhn2015}, wherein two sub-apertures of the 5.1~m Palomar telescope were injected (after phase delay) into one single-mode fibre. An additional benefit of a single-mode photonic waveguide approach is that the nulled output is already traveling in a single-mode fibre, making it convenient to feed into subsequent detectors or high-dispersion spectrograph.

In this paper we present the design and on-sky demonstration (at the 8~m Subaru telescope) of the next evolution of spatially-filtered nulling interferometry: the integrated photonic nuller. As opposed to using a single optical fiber, this technique, named GLINT (Guided-Light Interferometric Nulling Technology), injects separate sub-apertures (or, potentially, telescopes) into separate single-mode waveguides inscribed within a monolithic optical chip. The chip is manufactured using ultrafast laser inscription \citep{Nolte2003, Gattass2008, Arriola2013, Gross2015}, wherein a femtosecond laser is focused into the material to permanently modify the local refractive index, and then translating the material in three dimensions to precisely sculpt a system of single-mode waveguides with full three-dimensional freedom. The chip contains not only waveguides, but also splitters and directional couplers, to split and recombine the light (via evanescent coupling). Thus the actual interferometry takes place entirely within the chip, with just the intensity of the various output channels -- delivered to photodetectors via pigtailed fibres -- measured externally. A detailed description of the instrument, and the photonic circuitry, will be given in Section \ref{sec_InstDescription}. The application of this technique to high contrast imaging has previously been demonstrated in conventional (non-nulling) interferometry in the Dragonfly project \citep{Jovanovic2012, Norris2014} upon which this new instrument builds. The GLINT instrument presented here is closely linked to its sister project `GLINT South' \citep{Lagadec2018}, which is demonstrating the same photonic nulling technology on a non-AO corrected telescope (the 3.9~m Anglo-Australian Telescope in Australia).

The GLINT approach offers a number of advantages over previous single-mode nulling implementations. It simultaneously provides the bright (anti-null) output and  photometric channels along with the actual null channel, which allow more accurate estimate of the true null-depth to be realised (as opposed to non-simultaneous photometric measurements using chopping). More importantly, it is easily scalable to a larger number of input telescopes and baselines, with more complex arrangements of splitters, couplers and waveguides being straightforward to implement simply by adding features in the direct-write process. A multi-tier approach can also be implemented within the chip, wherein the nulled outputs from the first stage of couplers are coherently re-combined in additional stages of couplers. This would enable solutions that optimise the shape of the null \cite{Angel1997}, measure a nulling-interferometry analog of closure-phase \citep{Lacour2014} or perform kernel nulling, providing extra robustness against time-varying instrumental phase \citep{Martinache2018_arxiv}.

In the case where a photonic nuller combines multiple sub-apertures of a single telescope pupil (as opposed to between separate telescopes, as with long-baseline interferometry) some sort of \emph{pupil-remapping} is needed. This rearranges the two-dimensional array of sub-apertures into an appropriate configuration for the nulling chip to receive. Crucially, this needs to be done while precisely maintaining the optical path length between all sub-apertures, in order to maintain coherence. This can be done with separate optical fibres \citep{Huby2012} or directly within a three-dimensional photonic chip using direct write \citep{Jovanovic2012, Norris2014}. The latter method has the advantage that the optical path lengths of the different arms may be precisely matched during design despite their circuitous routes, by using design optimisation tools. The GLINT design implements waveguides as a series of three-dimensional Bezier curves, whose parameters are numerically optimised to match optical path length while maintaining the necessary waveguide separation, minimising total length and maximising bend radius \citep{Charles2012}. Moreover, since they are embedded within a single monolithic block, they are more robust against optical path delay differences due to temperature or mechanical vibration than bulk fibres. A key advantage of the GLINT nuller design is that, since both the remapper and interferometric portions are written using the direct write process, both functions can be combined into a single device.

In Section \ref{sec_InstDescription} the design of the current GLINT pathfinder instrument will be presented, including details of the photonic chip itself, the larger instrument and integration into the SCExAO extreme-AO system. In Section \ref{sec_DataAnalysis} the theory of the self-calibrating data analysis method will be described, and demonstrated with laboratory measurements. The results from the on-sky tests at the Subaru telescope will then be presented in Section \ref{sec_OnSky}. Our conclusions, including descriptions of the next steps in the GLINT instrument development, will be presented in Section \ref{sec_Conclusion}.

\section{The GLINT instrument: technical description}
\label{sec_InstDescription}
The GLINT pathfinder instrument was integrated and commissioned at the Subaru Telescope in March 2016. It was deployed as a module in the Subaru Coronagraphic Extreme Adaptive Optics system (SCExAO) \citep{Guyon2011, Jovanovic2013, Jovanovic2015}. The null is produced by injecting two circular sub-apertures extracted from either side of the Subaru Telescope pupil into the photonic chip, resulting in an effective baseline of 5.55~m. This is smaller than the maximum baseline length available from the 7.9~m Subaru Telescope pupil in the IR due to the practicalities of the reimaging system, and will be increased in future iterations. It operates at 1.6~$\mu$m with a bandwidth of 50~nm. This narrow band was used due to the chromatic nature of the $\sim \pi/2$ radian phase shift between waveguides (since it is currently produced by air delay) and directional coupler, and non spectrally-dispersed outputs. 

SCExAO is designed to produce a high Strehl ratio image optimal for small IWA coronagraphs, and hence acts as a `fringe-tracker' for GLINT, keeping the relative phase-delay between the two sub-apertures as constant as possible. The residual errors are self-calibrated using the statistical approach described in Section \ref{sec_DataAnalysis}. 

The coupling of each sub-aperture into its corresponding waveguide, and their differential phase, is maximised by a segmented deformable mirror internal to GLINT that controls the tip, tilt, and piston of both the sub-apertures independently. To perform measurements, first the waveguide-injection and phase-delay are optimised off-sky using the calibration lamp (a super-continuum source) to provide a starting point. The tip and tilt of each segment is scanned in a raster pattern and the output flux in the corresponding photometric channel measured, and a bicubic interpolation used to find the optimum value.  
Likewise, the optimum null is found by differentially scanning the optical path difference between the waveguides while measuring the null output, and the deepest null (corresponding to the white-light fringe) is identified. The scans are repeated on-sky using these as a starting point to expedite the process. Once on target and the AO loop closed, the instrument samples all outputs at 64 kilo-samples/second, interleaving dark frames (for amplifier bias subtraction) periodically.

\subsection{Instrument layout and SCExAO integration}
\label{sec_InstLayoutIntegration}
\begin{figure*}
	\includegraphics[width=\textwidth]{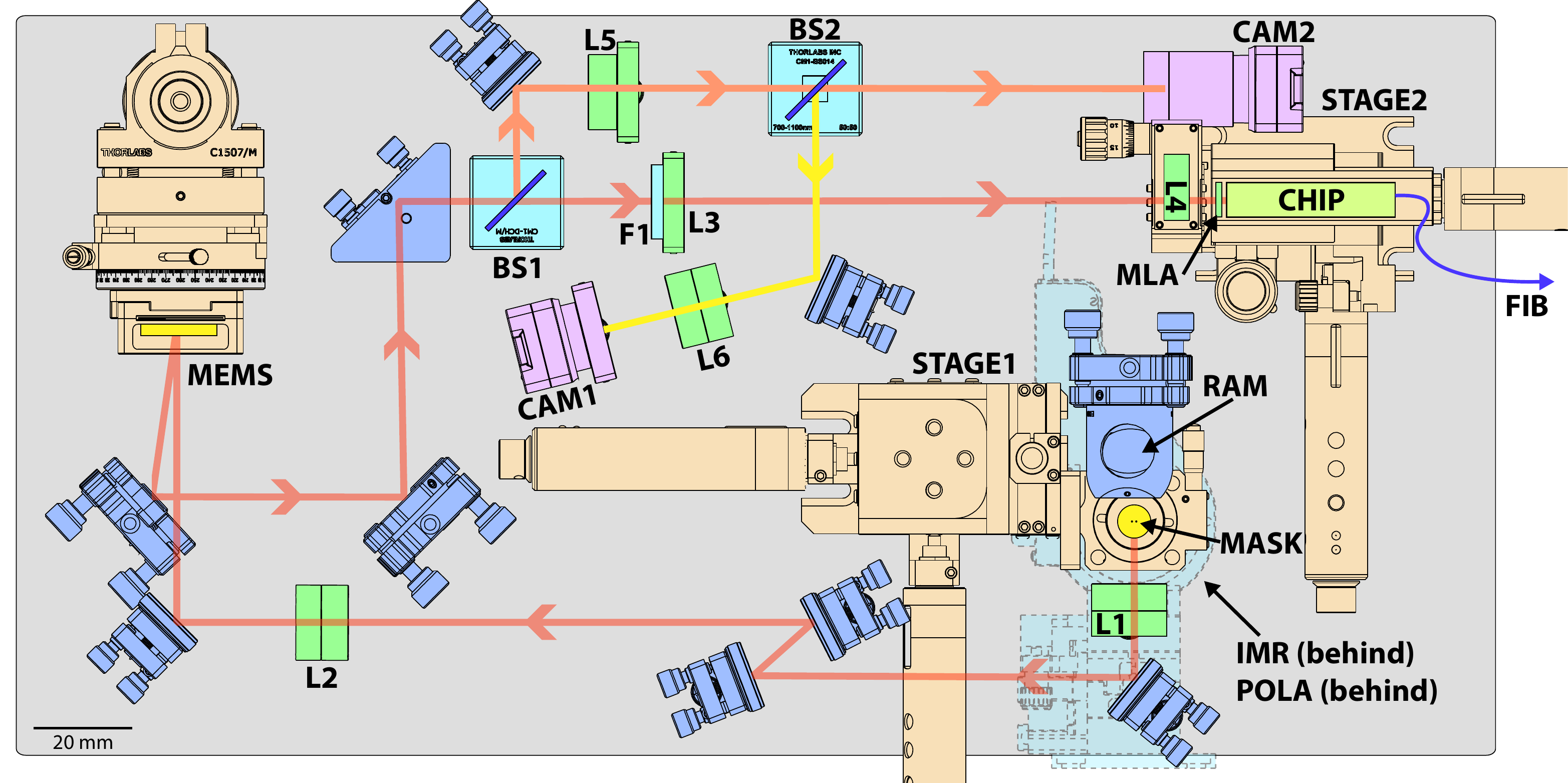}
    \caption{Schematic of the GLINT instrument layout. Light from SCExAO is steered via two Picomotor-driven mirrors (not shown) to pass through the linear polariser \textbf{(POLA)} and Image Rotator \textbf{(IMR)} (within a rotation stage), reaching the mask \textbf{(MASK)} which is at a pupil plane. The mask is mounted on a 2-axis Zaber actuated stage \textbf{(STAGE1)} for fine alignment. After being redirected via the right-angle mirror \textbf{(RAM)} (shown in withdrawn position) the pupil is reimaged via lenses \textbf{L1} and \textbf{L2} onto the MEMS segmented deformable mirror \textbf{(MEMS)}. A longpass dichroic beamsplitter \textbf{(BS1)} picks off light for the alignment cameras, and the pupil is reimaged by lens \textbf{L5} onto the pupil-viewing camera \textbf{(CAM2)}. A 50/50 beamsplitter \textbf{(BS2)} intercepts the beam before CAM2, and lens \textbf{L6} focuses the image plane onto the image-viewing camera \textbf{(CAM1)}.
Meanwhile, the beam transmitted through \textbf{BS1} passes through the bandpass filter \textbf{F1} and is reduced by beam reducing optics \textbf{L3} and \textbf{L4}, which also re-images the pupil onto the microlens array \textbf{(MLA)} at the front of the photonic unit. This then injects the sub-beams into the photonic chip \textbf{(CHIP)}, which is mounted on a 3-axis Zaber actuated stage \textbf{(STAGE2)} for precise alignment. Light from the 4 output waveguides is then transmitted via a fibre cable \textbf{(FIB)} to photodetectors (not shown).}
    \label{fig_InstSchematic}
\end{figure*}

The overall instrument design is as follows, with a schematic showing the major components and detailed beam path given in Figure~\ref{fig_InstSchematic}.

Starlight is first delivered to the AO188 facility AO system, where it receives lower order wavefront correction. It then reaches the near-IR optical table of SCExAO, where the IR light experiences high-order correction via a 2K actuator MEMS-based deformable mirror. The visible light component ($\lambda < 1 \mu$m) is split off and sent to a separate visible bench whereupon it is analysed by SCExAO's pyramid wavefront sensor to drive the high-order correction. Meanwhile the IR beam, which is usually sent to the IR science cameras, is redirected by a remotely-operable pickoff mirror and sent to the GLINT module via a focusing lens and a pair of dedicated steering mirrors (actuated by Newport Picomotor Piezo Mirror Mounts) to allow fine tuning of image and pupil position in the nuller. The beam then exits the near-IR optical table of SCExAO and enters the GLINT nuller optical table, which is mounted vertically on the side of the SCExAO support frame. An image rotator (labelled IMR in Figure \ref{fig_InstSchematic}) consisting of a dove prism and motor-driven rotation mount is used to rotate the image, enabling the Bracewell nulling mode. To aid in the characterisation of the polarisation dependence of the system, a Glan-Thompson linear polariser (POLA) can be inserted to enforce a single linear polarisation. 

The telescope pupil is re-imaged onto an opaque brass mask (MASK), containing two laser-cut holes corresponding to the desired sub-apertures. By translating the mask laterally (i.e. in the directions perpendicular to the beam) using a 2 axis stage (STAGE1), these holes are carefully aligned with the appropriate MEMS mirror segments and waveguides to prevent unused light entering the bulk of the chip and propagating, unguided, to the outputs. The pupil is again re-imaged (with lenses L1 and L2) onto the MEMS segmented deformable mirror. This mirror (model PTT111, manufactured by IrisAO) divides the pupil into 37 hexagonal segments, each of which has individual tip, tilt ($\sim \pm 3 mrad)$) and piston ($\sim \pm 3 \mu m)$) control. For the pathfinder instrument only two sub-apertures (and hence segments) are used, but potentially the entire telescope pupil can be used by increasing the number of waveguides. The MEMS mirror is used to optimise the injection into each waveguide and tune the phase delay between them.

Following the MEMS mirror, the pupil is re-imaged onto the microlens array (MLA) via lenses L3 and L4, which reduce the beam diameter from 4.2~mm to 210~$\mu$m (20:1 compression factor). Each lenslet has a diameter of 30~$\mu$m. These inject each sub-aperture into a corresponding waveguide on the end-face of the photonic chip. The MLA and chip are pre-aligned and focused in the laboratory and bonded in place, and then mounted in a single protective mount on a 3-axis stage (STAGE2). The 4 outputs of the chip (null, anti-null and 2 photometric outputs) are sent to photodetectors over an optical fibre cable (FIB). The entire instrument is designed for remote operation and alignment, with the mask, chip, steering mirrors and image rotator all using precision actuators, in addition to the MEMS. 

Diagnostics of both the image plane and pupil plane are obtained with two CMOS cameras, fed by a dichroic beam-splitter (BS1). The image-viewing camera (CAM1) allows the PSF position to be aligned to a known location, resulting in the beam appearing `face-on' to the chip and minimising the deflection needed in the MEMS segments when optimising injection efficiency. The pupil-viewing camera (CAM2) images the MEMS surface, the 2-hole mask and SCExAO's telescope spider mask upstream, with these 3 pupil planes appearing superimposed. This allows precise alignment of the mask with respect to the MEMS segments, and the whole instrument with respect to the SCExAO pupil (using the dedicated tip/tilt mirrors located within SCExAO). These are key to successful remote alignment. The pupil-viewing camera is also used in the chip coarse-alignment process; a requirement when the chip is replaced or the system has large misalignments. Here, the chip is back-illuminated by sending laser light through its output fibres. This projects two spots (one for each input waveguide) onto the 2-hole mask. By viewing these spots on the mask in the pupil-viewing camera, the chip can be translated in its two lateral directions to position the spots on the mask holes, and longitudinally to optimise focus by making the spots as small as possible.

The photodetectors used were Femto OE-200-IN2 InGaAs-based photoreceivers. The conversion gain was set to $10^{11}$ V/W, limiting the temporal bandwidth to about 1 kHz (-3 dB), with a noise equivalent power of approximately 6 $\mathrm{fW/\sqrt{Hz}}$. Signals were then acquired via a National Instruments USB-6212 DAQ. Data acquisition, MEMS actuator control, and optimisation was performed using a custom-written Matlab program.

\subsection{The photonic heart of GLINT}

\begin{figure}
	\includegraphics[width=\columnwidth]{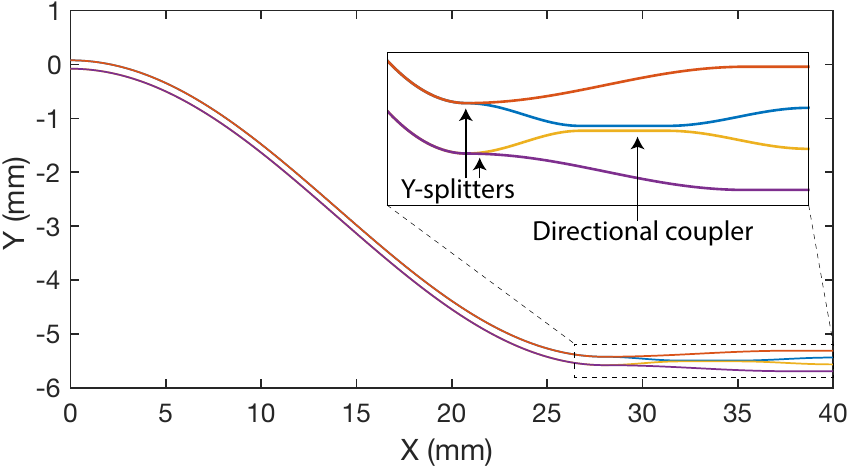}
    \caption{A top-down view of the waveguide arrangement in the photonic chip. The `side-step' feature (lateral displacement of the guides over the first 25~mm) avoids the effect of uncoupled light at the input interfering with guided light at the output. The inset shows a zoomed-in (horizontal 150\%, vertical 600\%) detail of the coupling region. Towards the left are the photometric splitters and in the centre is the evanescent coupler. The four output waveguides are butt-coupled to fibres at the output face.}
    \label{fig_ChipDiagram}
\end{figure}

The actual interferometry all takes place within the photonic chip. The chip employed in the pathfinder instrument has a relatively simple layout, illustrated in Figure \ref{fig_ChipDiagram}. Two waveguides, spaced by 155.9~$\mu$m in the horizontal plane originate at the input-face (on the left of the diagram). While the waveguide spacing at the chip-input can be arbitrarily chosen within manufacturing limits, this particular separation was chosen to match the exact pitch of a commercially available MLA (Suss Microoptics) at either side of the re-imaged telescope pupil. The first half of the chip consists of a large `side-step' formed by a cosine S-bend, where the waveguide position is translated laterally by 5.5~mm while maintaining matched path length. This is to avoid the effect of uncoupled light propagating unguided through the chip and interfering at the outputs, which has been shown to negatively impact measurement accuracy \citep{Norris2014}. The 28~mm long 'side-step' results in a minimum waveguide radius of curvature of 29~mm. In a chip with more than 2 inputs, this region is where the pupil-remapping would take place, converting the two-dimensional array of sub-pupils into a 1-dimensional (or other appropriate) array for subsequent beam combination. Next, each waveguide encounters a Y-splitter, where nominally 33\% of the light is split off and sent to separate photometric outputs. As will be shown in Section \ref{sec_DataAnalysis}, the simultaneous measurement of the coupled flux in each waveguide (time-varying due to seeing) is critical for accurate data analysis. Next, the waveguides form an evanescent directional coupler which was tuned to produce a 50-50 splitting ratio at its two output ports when co-phased coherent light enters the two inputs. Note that unlike in bulk-optic beam combination, where a $\pi$ radian delay must be introduced to shift from the central bright fringe to the adjacent dark fringe, in the case of a directional coupler a $\pi/2$ radian delay must be introduced. Correspondingly, if the light in either of the inputs is delayed by $\pi/2$ radians the coupler produces one entirely dark (null) channel and one bright channel. The coupler is created by bringing the two input waveguides together using cosine S-bends to a proximity of 10~$\mu m$ over an interaction length of 3.75~mm, prior to diverging again. These two outputs (the bright and null channels) then continue to the output. At the output face, the four waveguides are butt-coupled and permanently bonded using UV curing adhesive to a fibre V-groove with 127~$\mu m$ pitch, and sent via standard telecommunications single-mode fibres (SMF-28) to the photodetectors. The entire photonic chip measures 41~mm in length by approximately 10~mm in width, with a thickness of 0.7~mm. 

The single-mode waveguides were inscribed inside a monolithic block of boroaluminosilicate glass (Schott AF-45) using Ultrafast Laser Inscription, where a femtosecond pulsed laser is used to create a positive refractive index change inside the medium. The glass block was then translated using computer-controlled precision air-bearing stages allowing the laser to sculpt the desired waveguide circuitry in three dimensions. The 10.75~$\mu m$ diameter waveguides were written with a 800~nm wavelength, 5.1~MHz repetition rate Ti:sapphire laser at pulse energy of 45~nJ. The laser was focused 300~$\mu m$ below the top surface using a 100$\times$ 1.25~NA oil immersion microscope objective (Zeiss N-Achroplan) while the sample was translated at a velocity of 18.3~mm/s. The Y-junctions were formed by overpassing the side-step section twice, first to form the photometric branch of the Y-junction and in a second step to inscribe the arm of the directional coupler. 
\begin{figure}
	\includegraphics[width=\columnwidth]{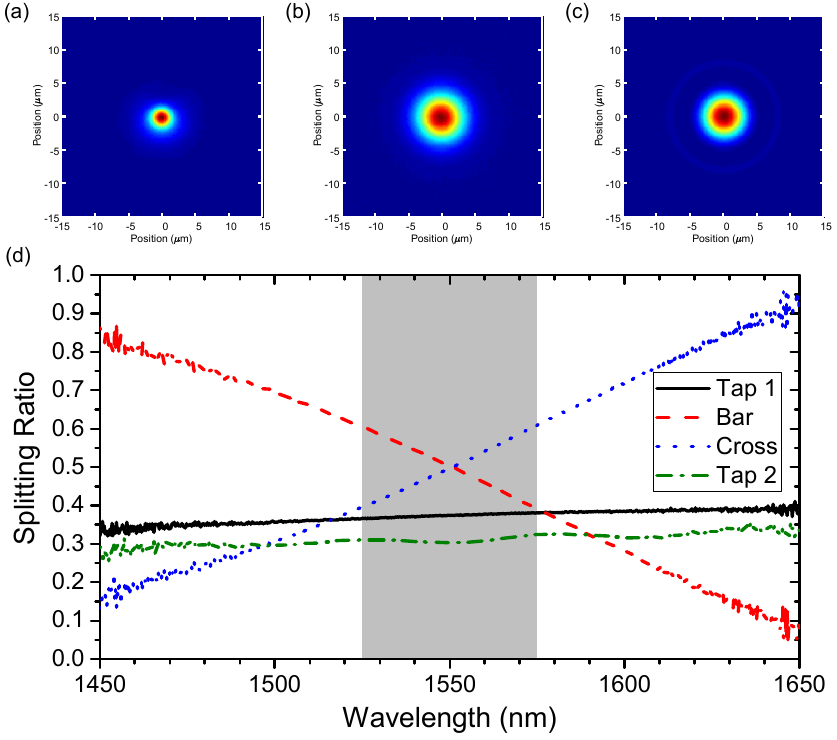}
    \caption{(a) Waveguide near-field profile at 1550~nm and for comparison (b) the measured near-field profile of standard telecommunication single-mode fibre (SMF-28e) and (c) the calculated focal spot of the MLA assuming the lenslets are diffraction limited. (d) Wavelength resolved splitting ratio of the directional coupler and both photometric taps. The red and blue lines show the power emerging from each of the two output waveguides of the coupler (labelled `bar' and `cross') when light is injected into one of its inputs. The grey shaded area indicates the transmission band of the 50~nm bandpass filter.}
    \label{fig_NFPsAndSplitting}
\end{figure}
The final single-mode waveguides had a mode-field profile with a 4$\sigma$ diameter of $9.3\times8.5$~$\mu m$ (H$\times$V), as shown in Fig.~\ref{fig_NFPsAndSplitting}(a). For comparison, Fig.~\ref{fig_NFPsAndSplitting}(b) shows the 11.0~$\mu m$ $4\sigma$ diameter near-field profile of standard telecommunications single-mode fibre, resulting in 7\% coupling loss between the chip and the fibre V-groove array. The measured splitting ratios of the Y-couplers (that provide the photometry channels) were $37\pm1\%$ and $31\pm1\%$, respectively, over the 50~nm operational wavelength window, see Fig.~\ref{fig_NFPsAndSplitting}(d). The directional coupler (for the nulling) exhibits nearly perfect 50-50 splitting at the central wavelength of 1550~nm, while changing to 40-60 and 60-40, respectively, at the edges of the operational wavelength band as indicated in  Fig.~\ref{fig_NFPsAndSplitting}(d).    

\begin{figure}
	\includegraphics[width=\columnwidth]{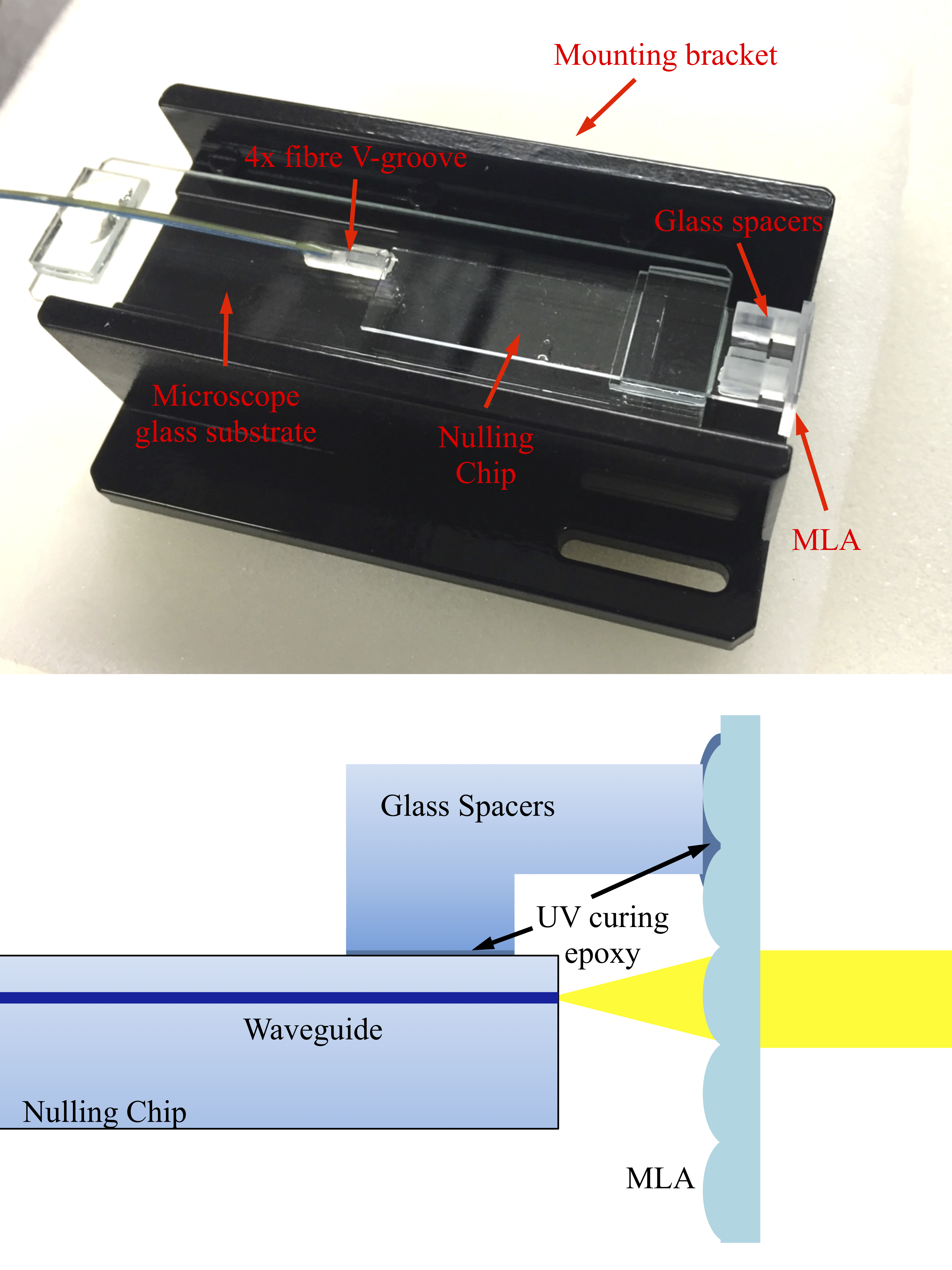}
    \caption{The photonic assembly includes the nulling chip, the MLA, and 4-port fibre v-groove, all aligned and bonded in the laboratory prior to being placed in a custom mounting bracket (top). The mounting bracket was attached to a precision translation stage inside GLINT such that the telescope pupil can be accurately placed at the correct spot on the MLA, with no further alignment of the photonics required. Looking edge-on at the entrance of the chip (bottom), a novel bonding technique was used to permanently bond the MLA to the chip, while maintaining accurate alignment. Custom 'L' shaped glass spacers were used to ensure that the UV curing epoxy bonded region did not contaminate the lenslets being use for focusing the light into the waveguides.}
    \label{fig_MLAbondingdiagram}
\end{figure}

The MLA features 30~$\mu m$ diameter lenslets on a hexagonal pitch of 30~$\mu m$ with a focal length of 96~$\mu m$ and a numerical aperture of 0.16. The fused silica substrate of the MLA was 10x10~mm in area with a 1~mm thickness. Because a specific NA was required to match the acceptance angle of the single-mode waveguides, the focal point of the lenslets happened to be inside the glass substrate. This forced the MLA to be oriented `backwards' with the flat substrate facing the collimated beam, and the convex side facing the photonic chip. To avoid thermal drift between the MLA and photonic chip in GLINT, the photonics and MLA were permanently bonded in a complete package as shown in the upper part of Figure \ref{fig_MLAbondingdiagram}. This also meant the considerable complexity of aligning these components only needed to be performed once, rather than during on-sky instrument alignment. Unfortunately, simply placing UV curing epoxy between the MLA and photonics would not work as this changes the refractive index of the glass-air interface required for the MLAs to function at specification. Thus a more complicated bonding method was used where two glass 'L' shaped spacers were attached to the top of the photonics and the front of the MLA chip (see bottom figure of \ref{fig_MLAbondingdiagram}). The alignment procedure was to first co-align the angle of the MLA and photonic chip end-face to a collimated reference beam using the Fresnel back-reflection off the MLA substrate and chip end-face, respectively. This is critical as any angular misalignment between the MLA and reference beam will result in a phase ramp across the input when the system is aligned for optimum injection efficiency, leading to a poorer null depth.  Once the MLA is perfectly face-on, the photonic chip (with output V-groove already attached) was brought into optimal alignment by being back-illuminated via the attached optical fibres. The distance between chip and MLA was adjusted by measuring the gap using a calibrated vision system. In the next step, the pupil of the MLA was imaged onto a detector and the light emerging from the two waveguides was centred with respect to the pupils of the lenslets by translating the chip using a high-precision, piezo actuated 6-axis translation stage. Once the initial alignment was optimised, the L-spacers were placed on top of the chip and attached with UV curing epoxy (Norland NOA61) on the mating faces. Once the MLA curing was complete, the entire MLA-photonic-V-groove assembly was attached to a microscope slide with further glass spacers to maximise rigidity of any overhanging parts and strain relief the optical fibres. Finally, the assembly is bonded to a custom mount plate to be placed in GLINT.

The internal transmission of the 41~mm long chip was measured to be $86\pm1\%$ at 1550~nm providing an upper bound for the waveguide propagation losses of $0.14\pm0.04$~dB/cm. The internal transmission includes losses due to propagation, bend losses and absorption caused by impurities of the substrate material ($\sim1\%$ \cite{Meany2014}). The coupling losses between the fibre array and waveguide chip are 7\%, found by calculating the mode-overlap integral between the measured waveguide and fibre near-field. Additional losses occur due the mismatch between MLA focal spot and waveguide mode. Assuming the MLA is diffraction limited, the maximum coupling efficiency is 68\% if the entire aperture of the lenslet is used. Furthermore, Fresnel reflection losses totalling 10\% occur at the uncoated surfaces of the MLA and chip input face. This limits the maximum throughput of the entire assembly to 49\% if every component is optimally aligned.

\section{Data analysis}
\label{sec_DataAnalysis} 
Here we adopt the formalism defined in \cite{Serabyn2000}. The key scientific observable produced by a nulling interferometer is (for each baseline) the \emph{null depth} ($N$), defined as
\begin{equation}
	N = \frac{I_-}{I_+}
\end{equation}
where $I_-$ and $I_+$ are the intensity of the destructive and constructive fringes respectively. 
The null-depth is closely related to the \emph{visibility} quantity in conventional interferometry, and effectively measures the coherence properties of the light, with $N > 0$ indicating some degree of incoherence. Of scientific interest is the measurement of the spatial coherence of the astrophysical target, which contains information about the spatial structure of the object. However the measured null depth also includes contributions from \emph{instrumental leakage} terms as well as from the astrophysical null, and in practice the instrumental leakage component may be much larger than the underlying astrophysical null $N_A$. 

As detailed below there are various factors determining the instrumental leakage, two of the most critical being the matching of the anti-phase condition (precise offset by $\pi/2$ radians) over the input waveguides, and having equal injection efficiency in both channels. These terms in particular are time-varying and impacted by the seeing. The extent to which both of these terms can be stabilised is set by the performance of the SCExAO adaptive optics system, however even under ideal conditions it is inevitable that these terms are significant. Rather than trying to build more complex and costly solutions to remove residual phase errors, we instead employed a statistical approach capable of handling non-ideal data, as pioneered by the Palomar Fiber Nuller group \citep{Hanot2011}. Essentially, rather than naively combining the measured quantities from a particular observation, we instead focus on the probability distribution function (PDF) of each measurable quantity, and compare it to a modelled set of PDFs. This method also has the advantage that it is self-calibrating, so observation of a separate unresolved calibrator star is not required.

In fact using this statistical approach proved to be particularly important as very large tip/tilt errors (due to telescope vibration) were encountered on-sky \citep{Lozi2016}. The resulting large differential phase errors also posed an extra numerical challenge. In the methods described by \cite{Hanot2011} it is assumed that the phase errors are small and so some small-number approximations can be used, described below. But in the case of the present observations these approximations could not be used and a full Monte-Carlo simulation of data had to be developed.

\subsection{Statistical analysis theory \& background}
\label{sec_analysisTheory}
Adopting the nomenclature used in \cite{Serabyn2000}, for coherent light of a single polarisation, the intensity of constructive ($I_+$) and destructive ($I_-$) interference fringes is 
\begin{equation}
	I_\pm = \frac{1}{2}( I_1 + I_2 \pm 2 \cos(\Delta\phi) \sqrt{I_1 I_2} )
	\label{eqn_Ipm}
\end{equation}
where $I_1$ and $I_2$ is the intensity of the input beams and $\Delta\phi = \phi_1 - \phi_2$ is the relative phase delay. The key observable in nulling interferometry is the \emph{null depth} $N$, defined as
\begin{equation}
\label{eqn_nulldepth}
	N = \frac{I_-}{I_+} .
\end{equation}
When fitting parameters, it is useful to define the \emph{mean intensity}:
\begin{equation}
	\left< I \right> = \frac{1}{2}(I_1 + I_2)
	\label{eqn_meanI}
\end{equation}
and the \emph{fractional deviation from mean intensity}:
\begin{equation}
	\delta I = \frac{I_1 - I_2}{2\left< I \right>}
\end{equation}
and then rewriting Equation \ref{eqn_Ipm} as
\begin{equation}
	I_\pm = \left< I \right> \left( 1 \pm \cos(\Delta\phi) \sqrt{1-(\delta I)^2} \right) .
	\label{eqn_IpmRearranged}
\end{equation}

When the magnitude of the phase error is small, some small-number approximations can be used to evaluate Equation \ref{eqn_IpmRearranged} directly \citep{Hanot2011}. But in our high phase and intensity error regime we can not do this, so must instead resort to a more complex and computationally expensive (Monte Carlo) method.

As per Equation \ref{eqn_nulldepth}, knowledge of the magnitude of both the destructive interference $N_-$ and constructive interference $N_+$ is needed. In a standard nulling interferometer (including fibre-based nullers) only $N_-$ is known, so $N_+$ must be estimated. With the aforementioned small phase-error approximations, $N_-$ can be assumed to be very small and so $N_+$ can simply be taken to be the same as the mean intensity (Equation \ref{eqn_meanI}). In our high phase-error case this is not appropriate, so we need a better estimate of $N_+$ (which we will denote $\widehat{I_+}$). Here we will test two different methods.

The first method is based on photometry. Although $\Delta\phi$ is not known, for GLINT the instantaneous power in each channel \emph{is} known, and so a $\widehat{I_+}$ estimate can be obtained by writing
\begin{equation}
	\widehat{I_+} = \frac{1}{2}( I_1 + I_2 \pm 2 \sqrt{I_1 I_2} )
	\label{eqn_IplusBetter}
\end{equation}
i.e. setting $\Delta \phi$ to zero. Despite the missing $\Delta \phi$ term, this was found to give a better estimate than simply using the mean intensity. For the second method, the output of the anti-null channel can be used directly as the value for $I_+$. But, as discussed in Section \ref{sec_LabResults}, this method performed poorly when large phase errors were present. 

To account for the deviation of our estimate from the true value, adopt the \emph{relative intensity deviation} parameter $I_r$ from \cite{Hanot2011}'s method, defined as
\begin{equation}
	I_r = \frac{I_+}{\widehat{I_+}}.
	\label{eqn_IrDef}
\end{equation}
This will be a free parameter in our fit.

The background term (in our case dominated by dark current) must also be accounted for. (The $I_1$, $I_2$ and $I_-$ terms written thus far implicitly assume some background bias has been subtracted off). While we know the instantaneous waveguide power thanks to the photometric channels, we do not know the instantaneous background value (i.e. because these are separate photodiodes). This introduces 4 new free parameters - the instantaneous background values for $I_1$, $I_2$, $I_+$ and $I_-$.

Lastly we have the astrophysical null term $N_A$. This is the observable quantity of interest, and describes the leakage arising from the astrophysical source. Note that this can actually be measured even if it is smaller than the instrumental null achieved. There is a straightforwards relationship between the visibility $V$ and the astrophysical null depth \citep{Mennesson2011}:
\begin{equation}
	N_A = \frac{1 - |V|}{1 + |V|}
\end{equation}

To generate a model, values for $\delta I$ and the backgrounds are required. The distributions for each of these are measured directly from the data via a histogram, and injected into the model (the so-called Numerical Self Calibration (NSC) method). This leaves 5 free parameters: $\Delta \phi_\mu$ and $\Delta \phi_\sigma$ (the mean and standard deviation of the relative phase delay); $I_{r_\mu}$ and $I_{r_\sigma}$ (the mean and standard deviation of the relative intensity deviation) and $N_A$ - the astrophysical null. It should be noted that while the inclusion of the PDFs of the other terms corresponds to a convolution of PDFs, and hence a change of shape of the final distribution, the astrophysical null term does not. Including the $N_A$ term is effectively a convolution by a delta function, and so corresponds to translating the entire PDF (to the right, as $N_A$ becomes positive).

A $\cos(\alpha)$ term describing the relative polarisation rotation between channels can also be included. Equation \ref{eqn_Ipm} assumes incident light of a single polarisation, matched between the two interferometric arms. Mixed polarisation states (such as a polarisation rotation or retardance) between the two arms will result in a shallower instrumental null. This would be a particular issue for long-baseline interferometry applications, where light destined for the different waveguides traverses a very different optical path, and may encounter different mirror angles, surfaces, etc. However in the present pupil remapping application, light for all waveguides traverses the same path, and to first order any systematic polarisation effects should be common between them, maintaining coherence. Therefore in this analysis, the $\cos(\alpha)$ term is neglected.

Chromatic effects are a major contributor to increased instrumental leakage. The above analysis assumes monochromatic light, however the actual instrument operates over a 50~nm bandwidth (centred at 1550~nm). Shallower null depth from broadband interference is not only due to the change in $\Delta\phi$ with $\lambda$ (i.e. it is not $\pi/2$ radians across the whole band) which is easily calculated, but also to the specific chromatic dependence on the coupling-ratios in the photonic devices. 
Based on laboratory measurements of the splitting ratio as a function of wavelength (e.g. as shown in Figure \ref{fig_NFPsAndSplitting}), the instrumental null-depth increase owing to chromatic effects is calculated to be $4\pm1\times 10^{-3}$. For the current analysis, this offset is subtracted from the measured null depths when stellar diameters are calculated from on-sky observations. For future iterations of the nulling chip, this coupling ratio will be optimally tuned in production and accurately measured after manufacture to allow precise calibration.

\subsection{Data analysis procedure}
The raw data obtained from an observation consists of a time-series of measurements of each of the 4 outputs (null, anti-null and two photometric channels), sampled several thousand times a second. Using the formalism described previously, for each moment in time the \emph{instantaneous} null-depth can be estimated from these 4 measurements. But this instantaneous value fluctuates greatly with time due to the seeing-induced phase and intensity variations, so we instead concern ourselves with the distribution of its values over the course of the observation (as well as the distributions of the individual outputs).

The basic approach is to estimate the probability distribution function (PDF) of the observed data by constructing a histogram, and then fitting a model PDF to this in order to constrain the free parameters. The model is generated via a Monte Carlo approach as described in Section \ref{sec_analysisTheory}. 

During observations, the control software automatically interleaves observations with dark measurements every few minutes. This allows the distribution of dark noise (both thermal and RF) to be measured, which may drift over time due to temperature change. 
To reduce the effect of dark current, the 64 kS/s sampling is binned down to 640 S/s, which is roughly consistent with the coherence time of the corrected seeing. This also allows an uncertainty for each binned data point to be estimated, which is the standard error in the mean of samples in that bin. Bias values, from the mean of the relevant dark measurements, are subtracted. This yields the time series of the null channel $I_-(t)$, anti-null channel $I_+(t)$ and the photometric channels $I_{1,2}(t)$.

Next, chip and fibre throughput, photodiode sensitivity and transimpedance amplifier gain differences are accounted for. These coefficients are measured in situ, before observations, by translating the pupil mask such that a light source is injected into one waveguide at a time (such that only incoherent transmission is measured). These are used to calculate normalised coefficients, which are then divided out in data reduction. $\widehat{I_+}(t)$, an estimate of $I_+(t)$, can then be obtained, using one of the two methods described in Section \ref{sec_analysisTheory} (i.e. either using the photometric outputs or using the anti-null output).

The estimated instantaneous null depth can then be calculated using
\begin{equation}
	\widehat{N}(t) = \frac{I_-(t)}{\widehat{I_+}(t)}.
	\label{eqn_Ntcalc}
\end{equation}

In the case of an ASC analysis, the next step is to measure the statistical properties which will directly constrain $\delta I$, and the background noise. $\delta I_\mu$ and $\delta I_\sigma$ are directly calculated as
\begin{equation}
	\delta I_\mu = \mathrm{mean}\left(\frac{I_1(t) - I_2(t)}{I_1(t) + I_2(t)}\right) , \qquad \delta I_\sigma = \mathrm{std}\left(\frac{I_1(t) - I_2(t)}{I_1(t) + I_2(t)}\right).
\end{equation}
$B_{\sigma I_1}$, $B_{\sigma I_2}$, $B_{\sigma I_-}$ and $B_{\sigma I_+}$ are calculated by taking the standard deviation of the dark measurements for the appropriate channel.

Finally the PDF can be estimated by way of a histogram. A key requirement is that measurement errors (originally derived from the SEM of a bin) propagate through to the PDF estimate, and also that these uncertainties on the PDF are cognisant of the number and distribution of values in a bin (this may be particularly important in the photon-noise-limited regime, where data nearer the null (i.e. with less stellar photon noise) should more strongly constrain the fit). The approach taken is to calculate the probability that a given data point $x_i$ with associated uncertainty $\sigma_i$ comes from a given bin. Thus the number of observations in bin $k$ is the sum of Bernoulli random variables, with the probability of each ($p_i(k)$) being the proportion of a normal distribution (with mean $x_i$ and standard deviation $\sigma_i$) that lies within the bin $j$. That is,
\begin{equation}
	p_i(j) = \int_{l_j}^{u_j} \frac{1}{\sqrt{2 \pi \sigma_i}} e^{{-\frac{(x_i - x)^2}{2\sigma_i^2}}} dz
\end{equation}
where $l_j$ and $u_j$ are the lower and upper limits of bin $j$ respectively. This can simply be calculated as the difference of two normal cumulative distribution functions. Then the uncertainty (standard error) of the bin can be calculated as
\begin{equation}
	\sum_{i=1}^n p_i(j) (1-p_i(j)) .
\end{equation}

To then generate a candidate model, vectors of random samples are created for $I_1$, $I_2$, $\Delta \phi$, $I_r$ and the four backgrounds $B_{I_1}$, $B_{I_2}$, $B_{I_-}$ and $B_{I_+}$. $\Delta \phi$ and $I_r$ are drawn from normal distributions $N(\Delta \phi_\mu, \Delta \phi_\sigma)$ and $N(I_{r_\mu}, I_{r_\sigma})$ (where the $\mu$ and $\sigma$ values are free model parameters). Samples for the remaining quantities are drawn from their measured distributions.  
Samples for $I_-$ can then be constructed using these samples as per Equation \ref{eqn_Ipm}. Samples for $\widehat{I_+}$ are calculated from these samples in accordance with the chosen $I_+$ estimation method (i.e. either from photometry or from the anti-null output). 
Finally, samples for $N$ can be calculated as per Equation \ref{eqn_Ntcalc}, and their PDF measured via a histogram, and fitted to the PDF of the data. 

The set of random samples for each term needs to be large enough that the error in the resulting PDF is very small. Empirically it was found that around $10^9$ samples are needed before results of subsequent trials are consistent (several times less than experimental error on the actual data). To make this fast enough for model-fitting, this was implemented on a GPU (an NVIDIA GTX 1080 Ti) using Matlab. Both random number generation (using the Philox4x32-10 algorithm) and vector arithmetic took place on the GPU, allowing a model of $10^9$ samples to be created in $\sim 3.5$~s. 

The fitting algorithm used is a trust-region-reflective non-linear least squares method. To calculate parameter uncertainties, a separate procedure is followed after the minimum is found, in which the $\Delta\chi^2 = 1$ contour is found by varying each parameter individually. 
Since this is a local optimiser, a basin hopping approach was used to find the global minimum, wherein the fit was run multiple times with random perturbations applied to the starting parameters and step sizes.

\subsection{Laboratory Measurements} 
\label{sec_LabResults}

\begin{figure*}
	\centering
	\subfloat[NSC, peak estimate from photometry]{
        \includegraphics[width=0.5\textwidth]{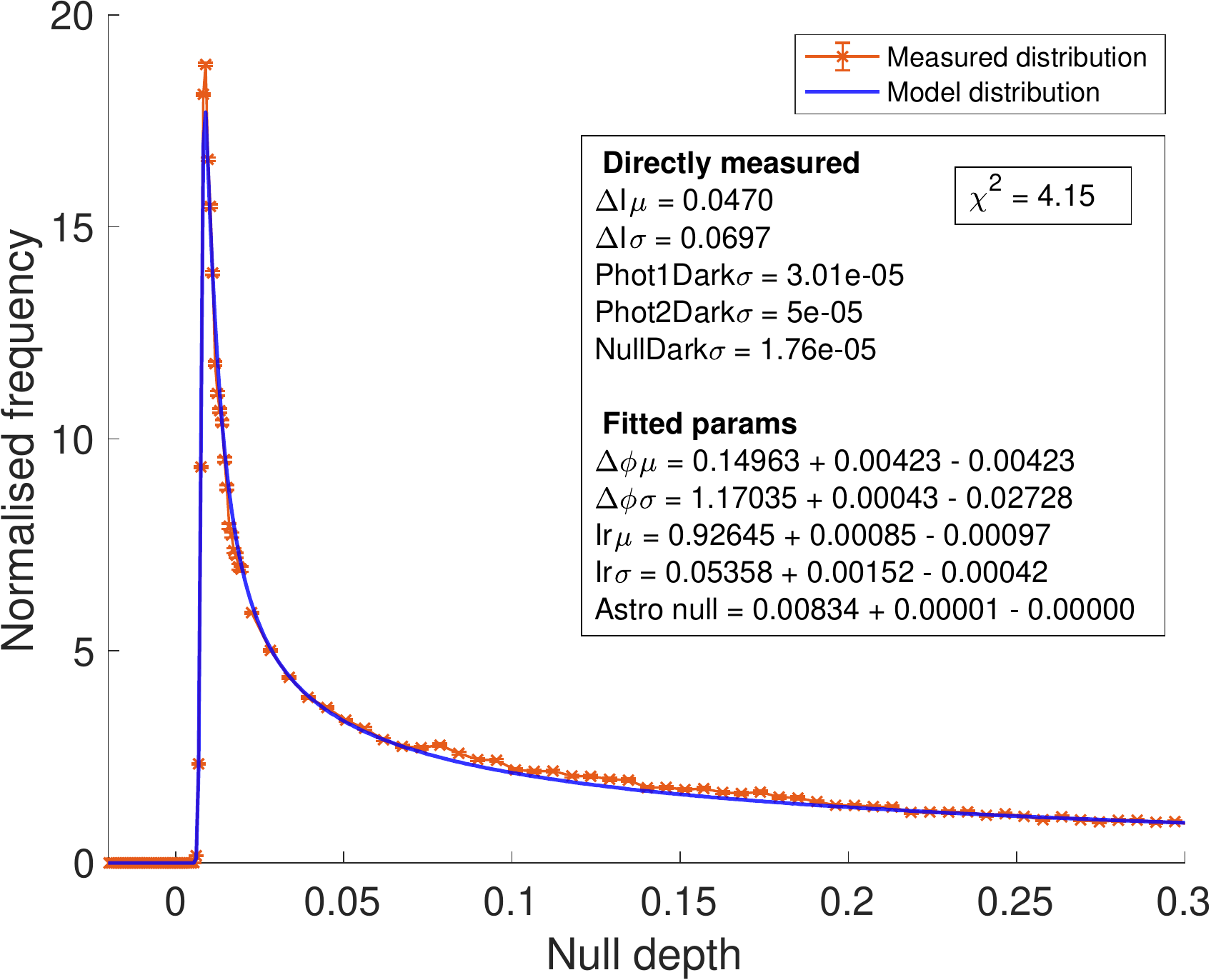}
        \label{}
    }
    \subfloat[NSC, peak estimate from anti-null channel]{
        \includegraphics[width=0.5\textwidth]{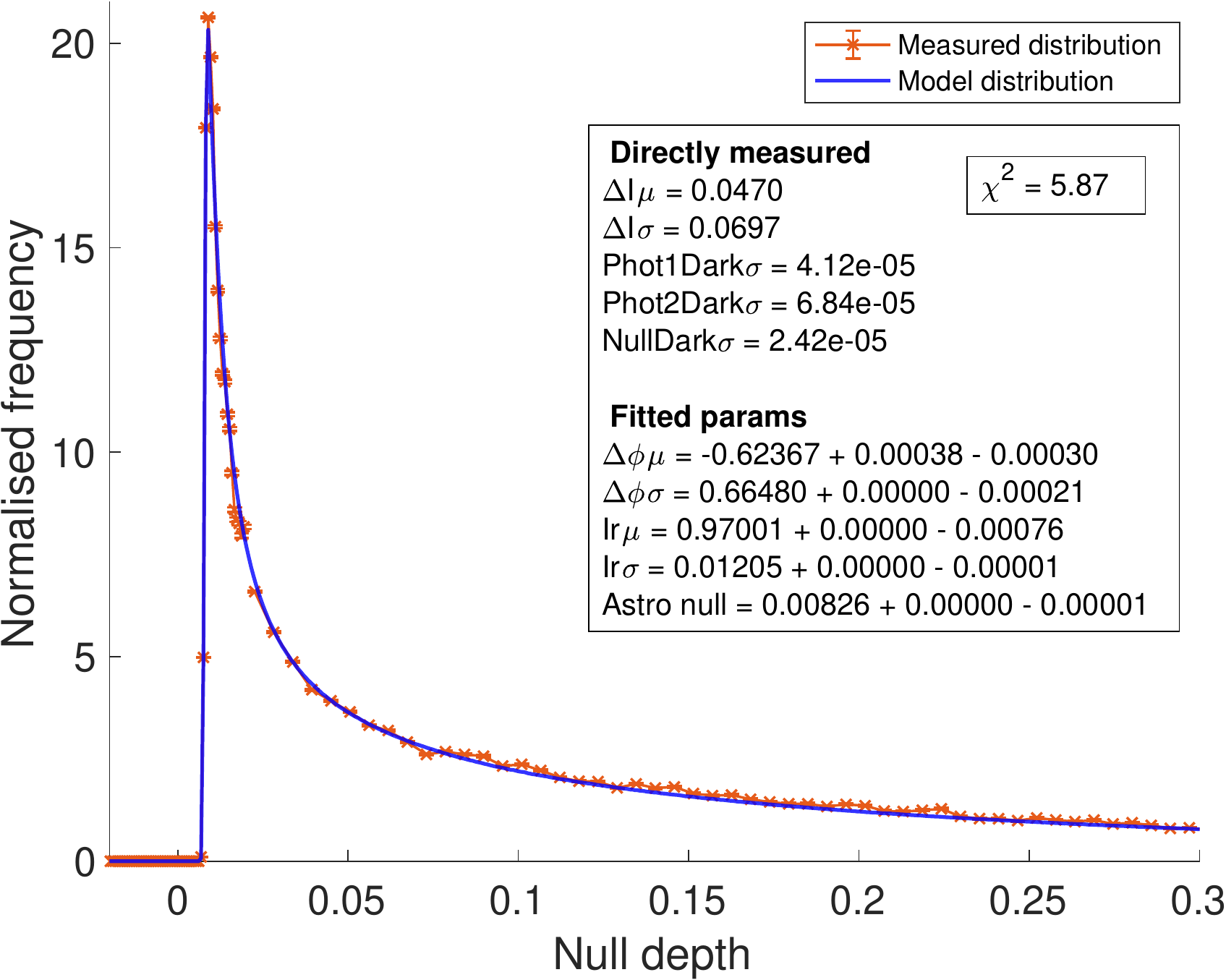}
        \label{}
    }
    \caption{Histograms and fitted model PDFs of NSC null-depth measurements from in-situ SCExAO off-sky tests, with $\sim$200~nm RMS simulated turbulence for 2 different analysis and fitting modes.  
In panel (a) the peak (or anti-null) is estimated from the photometry channels as per Equation \ref{eqn_IplusBetter}. Panel (b) instead uses the output from the `anti-null' channel of the chip to directly determine the peak estimate.  
Both peak estimate modes provide similar results, with a difference between their fitted nulls of 0.0001. This similarity was common to all measurements made with low phase error and low dark noise. Systematic `wiggles' in the PDF are seen, believed to be artefacts of the actuator spacing and discrete time steps of the deformable mirror. To improve clarity, data points at N>0.02 have been further binned for plotting.
    }
    \label{fig_results_LabScexaoMar2016_200nmWF}
\end{figure*}

Several laboratory measurements were made before on-sky observations, using the internal light source of SCExAO and its deformable mirror to simulate seeing by applying a moving Kolmogorov phase screen \cite{Jovanovic2015}. Of key importance is a comparison of the two different methods of estimating $I_+$ -- either using the photometric channels are the anti-null output -- as described in Section \ref{sec_analysisTheory}. Results from such a measurement are shown in Figure \ref{fig_results_LabScexaoMar2016_200nmWF}. Here, simulated turbulence with an amplitude of $\sim$200~nm RMS was added. The null is clearly seen in the histograms as a strong peak, and broadly fits with the model. The observed misfit arises from systematic `wiggles' seen in the PDF, which are believed to be caused by the temporal and spatial granularity of the phase pattern applied to the DM, and are not seen on-sky. The left panel shows the model and data using the photometry-based $N_+$ estimation method, and the right panel show sit using the anti-null output method. In this bright, low phase-error regime the results are consistent to within $10^{-4}$. The quoted uncertainty on the fitted $N_A$ of $\leq 10^{-5}$ is the statistical error on the fitted parameter, and does not include systematic errors.

As noted previously, the finite optical bandwidth of the measurement will lead to instrumental nulls > 0, since the analysis assumes monochromatic light. The expected null-depth increase arising from chromatic effects was calculated based on laboratory measurements of the wavelength-dependent coupling ratio of the chip to be $4\pm1\times 10^{-3}$ (limited by the accuracy of wavelength-dependent throughput measurements).
Accordingly, this contribution was subtracted from the measured null depths when stellar diameters are determined in Section \ref{sec_OnSky}. For the lab measurement it is noted that the instrumental null depth is around $4\times10^{-3}$ higher than that predicted from chromatic effects alone. This may be a result of the imperfect model fit, or due to fast temporal effects (e.g. simulated seeing and optical bench vibrations at speeds higher than our sample rate).

\section{On-sky measurements}
\label{sec_OnSky}
A number of resolved and barely-resolved stars were observed using the Subaru Telescope. In this section, the results of a range of on sky observations are presented in order to validate the instrument and demonstrate several different important effects. Firstly, successful operation of the instrument was demonstrated, with the angular diameter of barely-resolved stars measured using null-depths obtained via the numerical self-calibration approach. The two anti-null estimation methods are then compared and their goodness of fits examined. The surprisingly large phase errors found on sky are demonstrated, and the effect on null distribution and mitigation of their effects via model fitting are examined. This is compared to cases where these large phase errors are reduced by elimination of telescope vibrations using the adaptive optics system. Finally, the contribution from dark-noise, and particularly the effect of non-Gaussian dark noise distributions is evaluated.

The relationship between the limb-darkened diameter of a star and the astrophysical null depth \citep{Absil2006, Absil2011} can be described as:
\begin{equation}
	N_A = \left(\frac{\pi B \theta_\mathrm{LD}}{4 \lambda}\right)^2 \left(1 - \frac{7 u_\lambda}{15}\right) \left(1 - \frac{u_\lambda}{3}\right)^{-1}
	\label{eqn_diam}
\end{equation}
where $\theta_\mathrm{LD}$ is the limb-darkened stellar diameter, $u$ is the limb darkening coefficient, $\lambda$ is the centre observing wavelength and $B$ is the baseline length (for a uniform disk model set $u = 0$). For these observations, the uniform-disk (UD) diameters were used as the reference point, so $u$ was set to 0. 

\begin{figure*}
	\centering
	\subfloat[NSC, peak estimate from photometry, 5 parameter fit]{
        \includegraphics[width=0.5\textwidth]{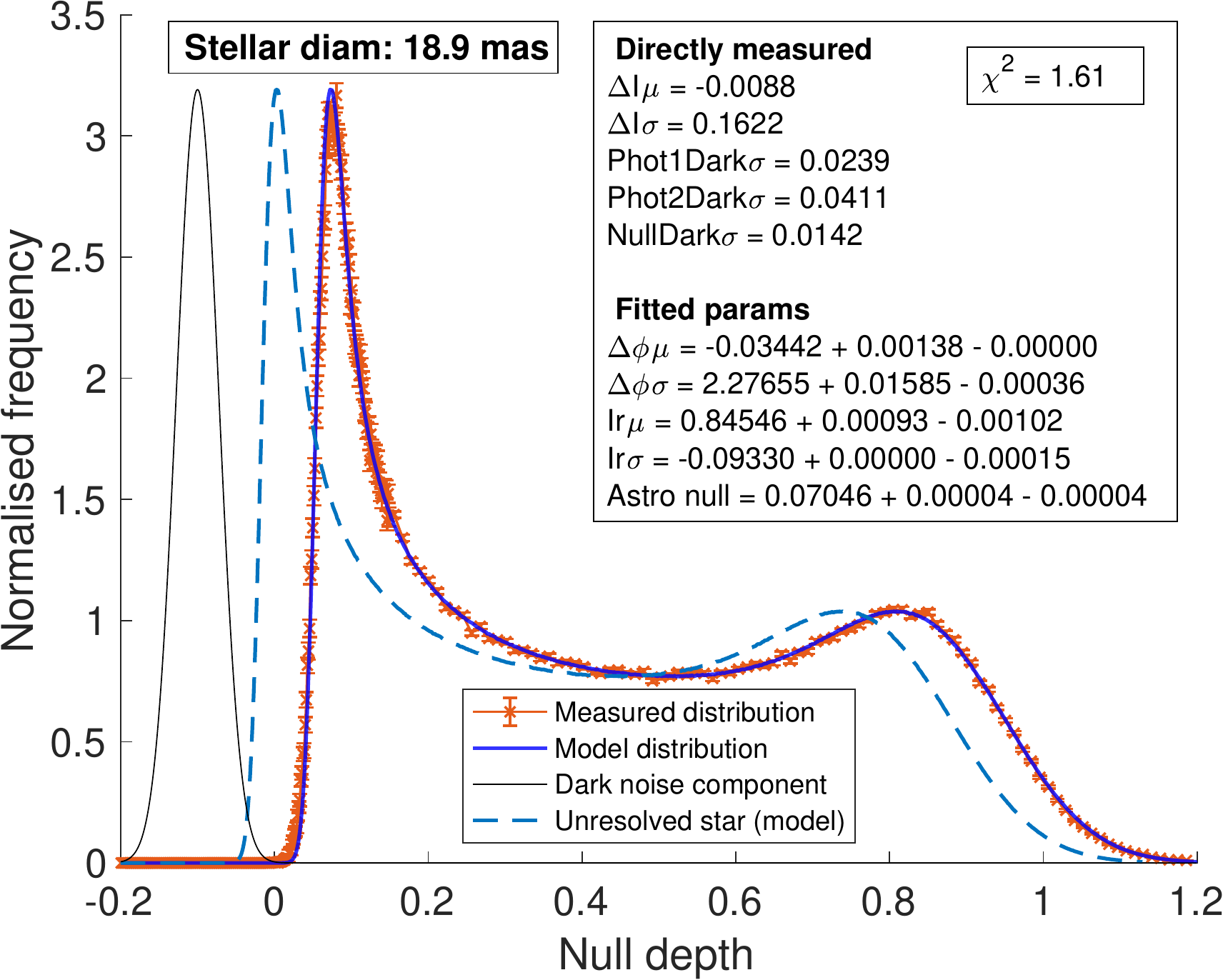}
        \label{}
    }
    \subfloat[NSC, peak estimate from anti-null channel, 5 parameter fit]{
        \includegraphics[width=0.5\textwidth]{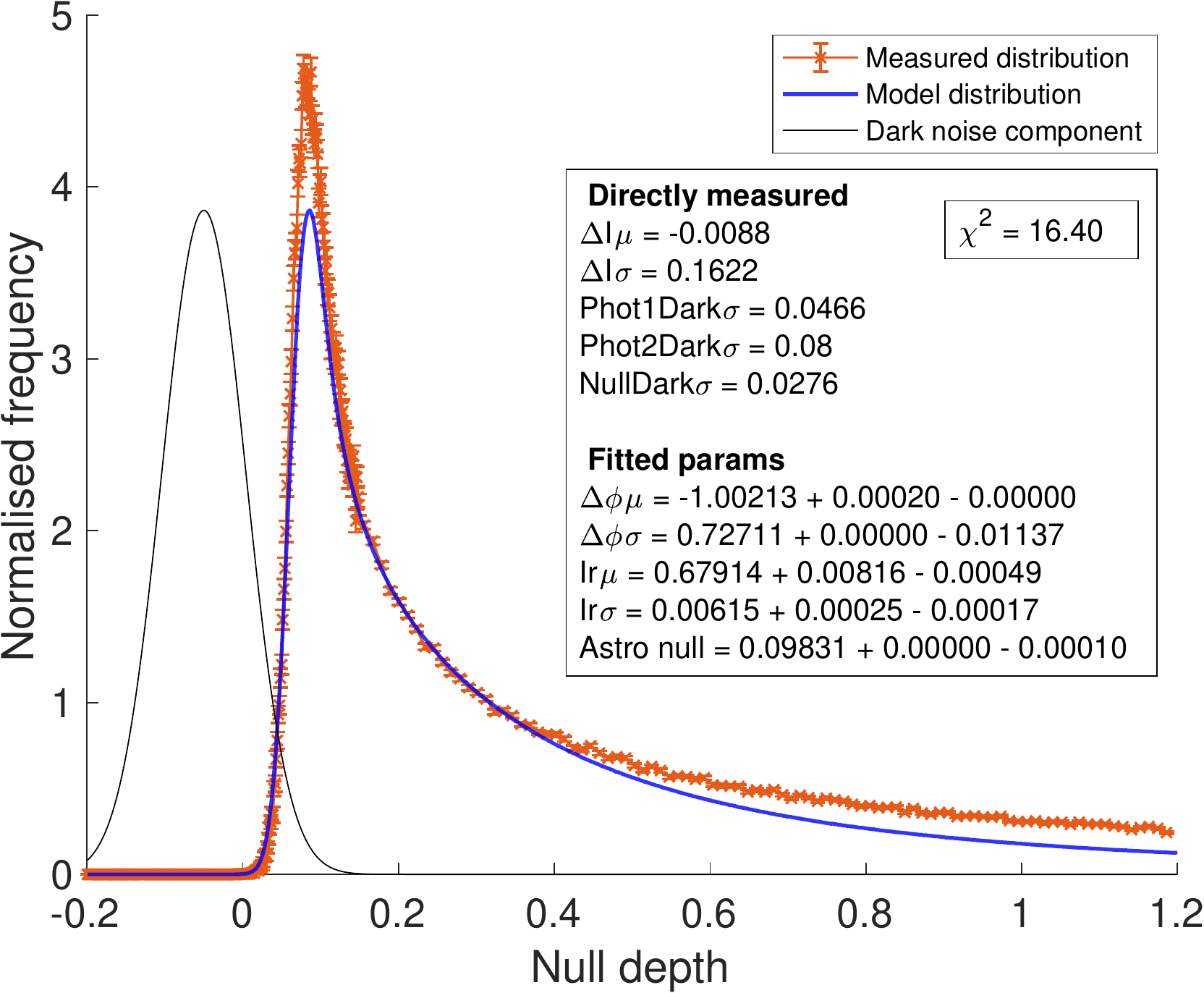}
        \label{}
    }
    \caption{ 
Histograms and fitted model PDFs of NSC null-depth measurements from 19 March 2016 observations of the K star $\alpha$ Bootis. The star is resolved by the GLINT Nuller, with an angular diameter of 18.9 mas measured from the self-calibrated null-depth. This is consistent with known values determined via interferometry. Note that the star is several times smaller than the formal diffraction limit. The model provides an excellent fit when using a peak estimate derived from the photometric channels (left panel), and a poor fit when using the anti-null channel (right panel) -- consistently seen when large phase errors are present.  
To improve clarity, data points at N>0.15 have been further binned for plotting. 
For reference, an unresolved star (using the same model) is over-plotted (blue broken line), clearly demonstrating that the star is resolved.
    }
    \label{fig_results_alfBoo201603}
\end{figure*}

Figure \ref{fig_results_alfBoo201603} shows the results from the March 2016 observations of the K star $\alpha$ Bootis. The measured astrophysical null is $0.0705 \pm 0.0004$ (statistical error). With a UD diameter of around 20~mas, this star has an apparent size only half the formal telescope diffraction limit ($\sim$ 50 mas at 1.6 $\mu$m), or indeed three times smaller considering the effective baseline length used here (5.5~m, corresponding to 73~mas). Despite this, the diameter was measured successfully, yielding a UD diameter (using Equation \ref{eqn_diam}) of 18.9 mas. This is consistent with known values determined by long-baseline interferometry (which has provided measurements ranging between 19.1 and 20.4~mas in K band \cite{Richichi2005}).

This observation also allowed the two $N_+$ estimation methods to be compared in the high phase-error regime. It is seen that when a peak estimate derived from the photometric channels is used, the model gives an excellent fit. However when the anti-null channel is used, the fit is poor. This is consistent with the observation that data with large $\Delta \phi _\sigma$ do not fit the model well when this peak estimate method is used. For the remainder of this Section, $N_+$ will be estimated using the photometry method. As a point of comparison, the same fitted model is over-plotted but with its astrophysical null set to zero, corresponding to an unresolved star. 

\begin{figure}
  \centering
  \includegraphics[width=0.5\textwidth]{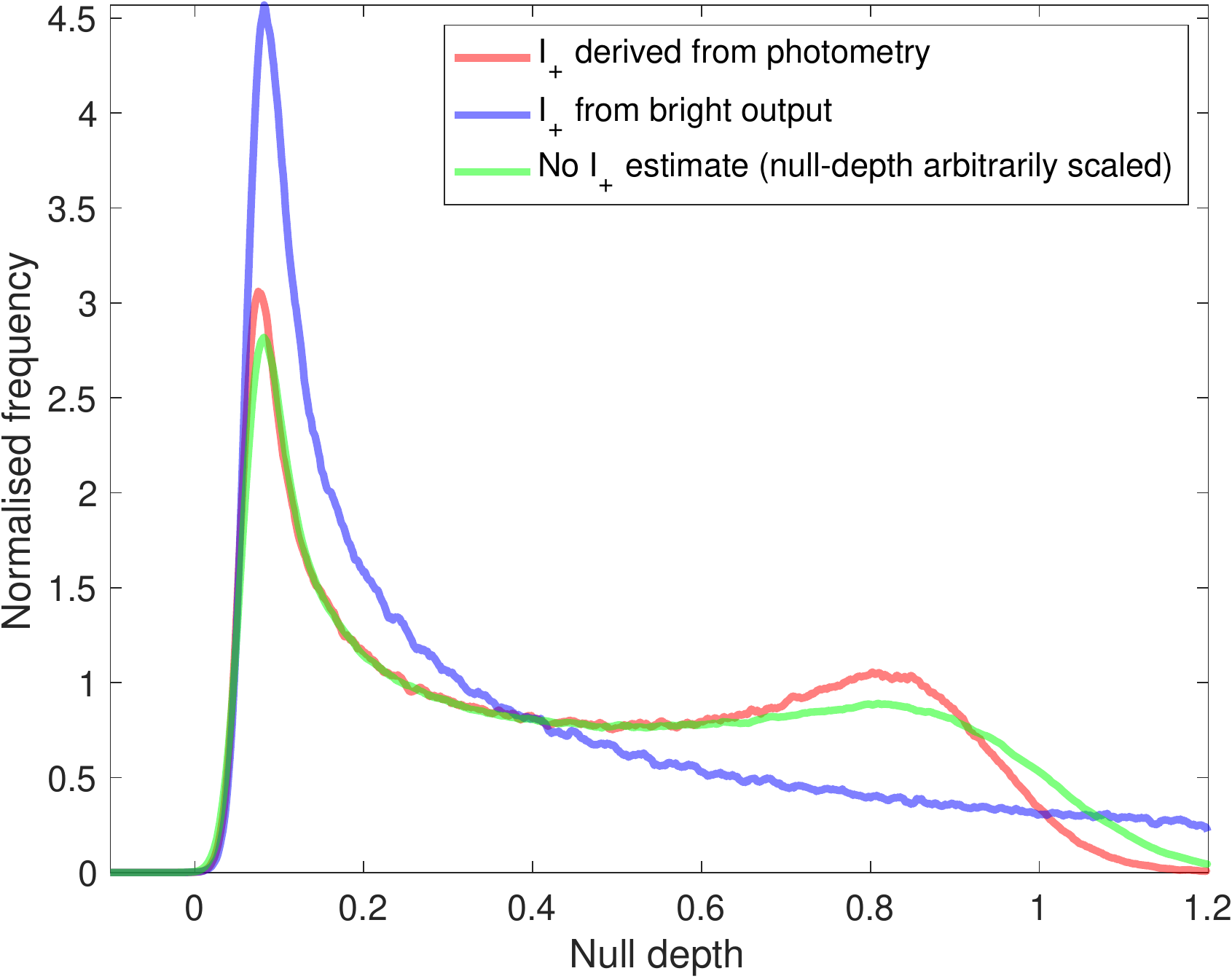}
  \caption{The histogram for the null-depth estimate $\hat{N}$ for the 2016 observations of $\alpha$ Bootis, derived by estimating $I_+$ using the photometric outputs (red), using the `bright' output (blue), as well as the \emph{raw} `null' channel output ($I_-$) (green), for an on-sky observation with large phase error. 
  See text for details.}
  \label{fig_comparePeakest}
\end{figure}

In this and many other cases the on-sky observations exhibited very large phase errors, exacerbated by a previously identified telescope vibration problem (see \cite{Lozi2018}). Furthermore dark noise tended to be quite high, due to the uncooled photodiodes and amplifiers used. Figure \ref{fig_comparePeakest} shows histograms of the null depth of the observation of $\alpha$ Bootis (with large phase errors) calculated using the two different methods of $I_+(t)$ estimation. It also includes the raw measurement of the output of the null channel $I_-(t)$ (which, since it is not-normalised by $I_+(t)$, is plotted in arbitrary units, and arbitrarily positioned horizontally). A distinctive `double hump', due to large phase errors of magnitude $\pi$ (and hence constructive interference occurring in the `null' channel) is clearly seen in the raw $I_-$ measurement. When the null depth is determined using the photometry-derived $I_+(t)$ estimate the double hump remains, with the second hump becoming more symmetrical and pointed. Since poor injection is correlated with large phase errors, this effect is likely due to correlated fluctuations in injection efficiency being largely divided out. Since this $I_+(t)$ estimate has no knowledge of the instantaneous $\Delta\phi$ this estimate is not precise (as demonstrated by the existence of the 2nd hump, which should not exist under the formal definition of null depth). However this is acceptable as the deviation of this estimate from the true value of $I_+(t)$ is fitted by two free parameters ($I_{r_\mu}$ and $I_{r_\sigma}$), as per Equation \ref{eqn_IrDef}.

In the case where $I_+(t)$ is taken to be the simultaneous measurement of the anti-null channel, the null depth derived is the `actual' null depth as defined by Equation \ref{eqn_nulldepth}. Ignoring intensity differences, the null depth here has the functional form $\frac{1 - \cos(\Delta\phi)}{1 + \cos(\Delta\phi)}$, which asymptotes to zero at large $\Delta\phi$.

\begin{figure}
  \centering
  \includegraphics[width=0.5\textwidth]{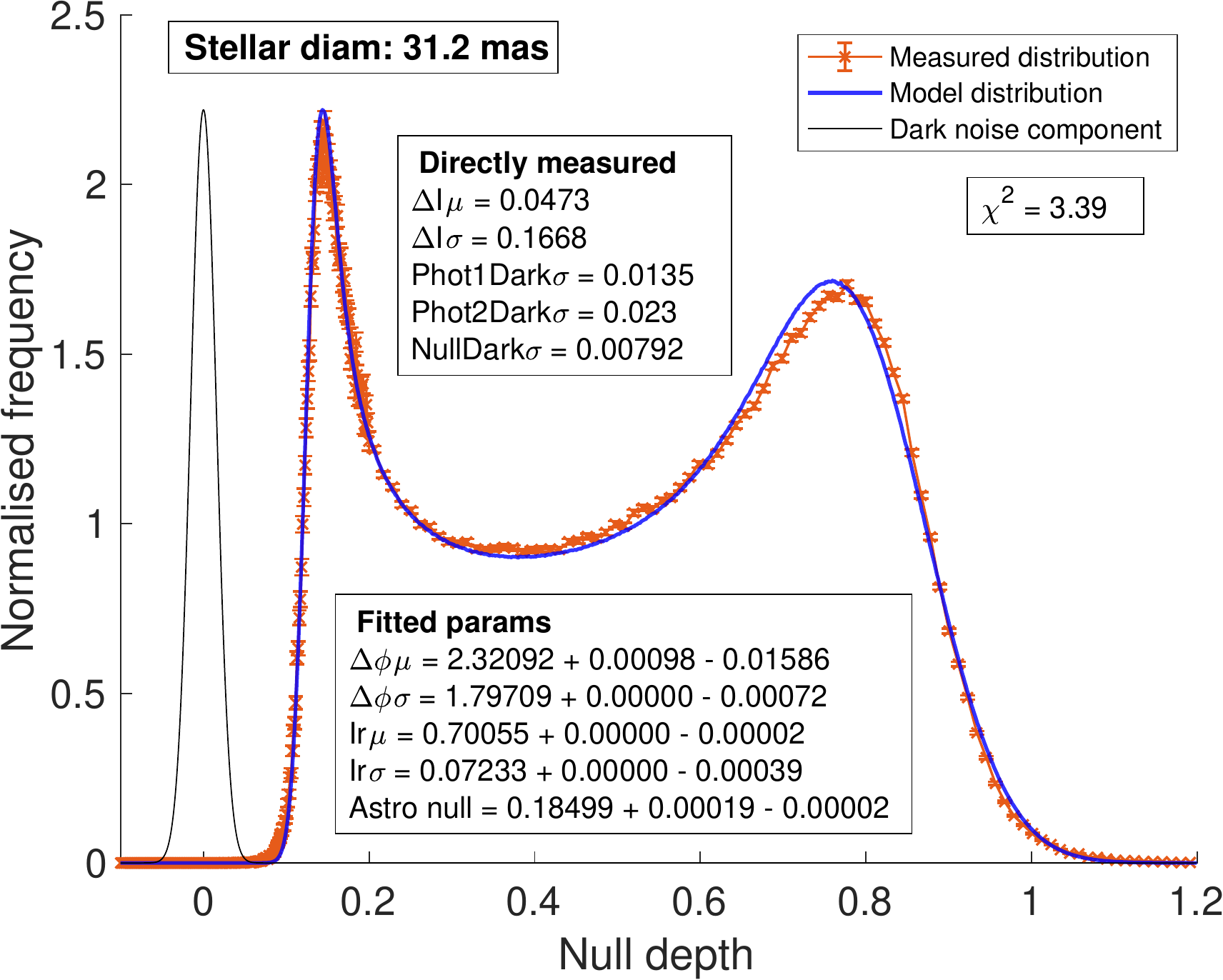}
  \caption{
  Histogram and fitted model PDFs of NSC null-depth measurements from 21 March 2016 observations of the red giant star $\alpha$ Herculis. The star is resolved by the GLINT Nuller, with an angular diameter of 31.2 mas measured from the self-calibrated null-depth. This is consistent with known values determined via interferometry. To improve clarity, data points at N>0.2 have been further binned for plotting.
}
  \label{fig_results_alfHer201603}
\end{figure} 

\begin{figure}
  \centering
  \includegraphics[width=0.5\textwidth]{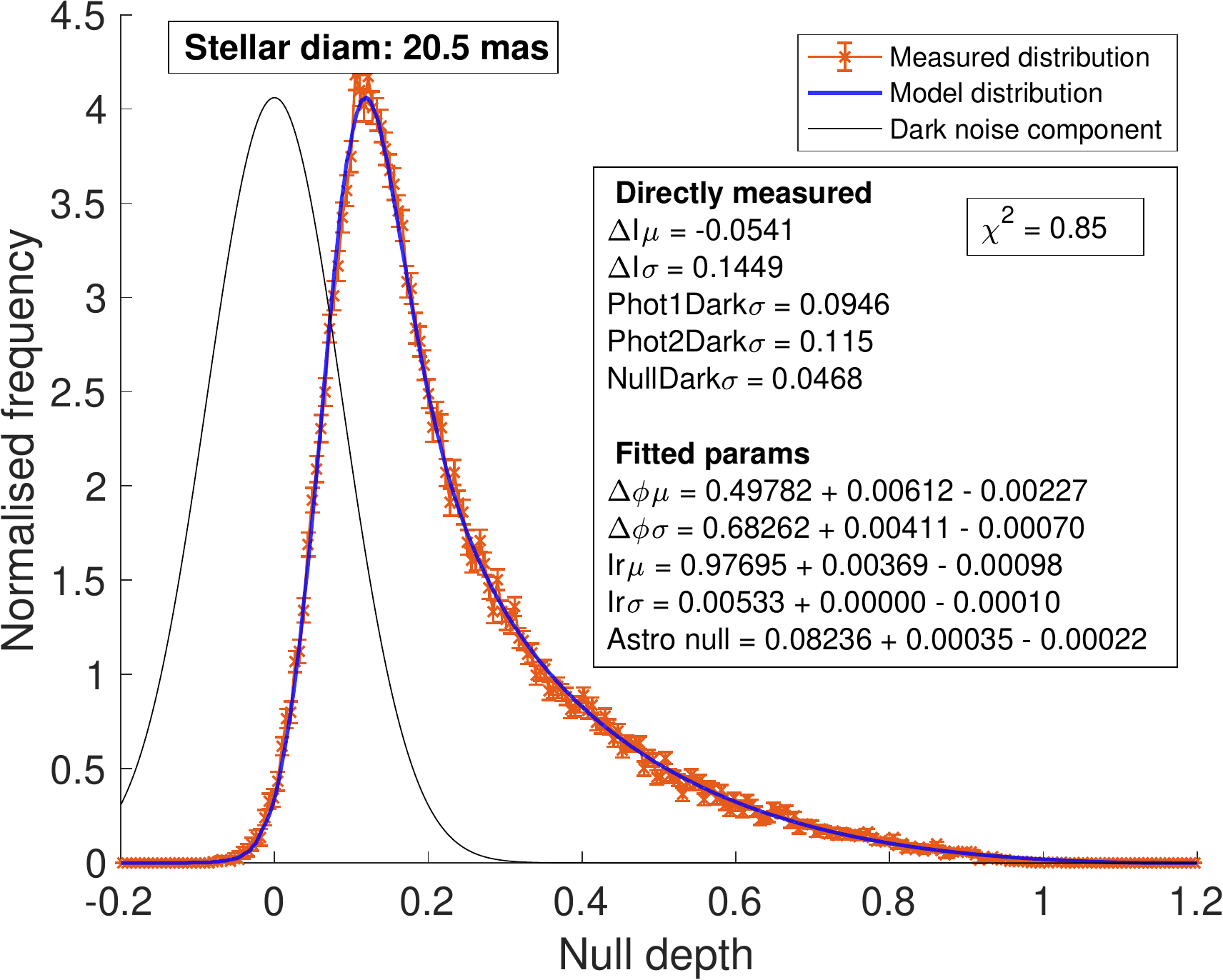}
  \caption{
  Histogram and fitted model PDFs of NSC null-depth measurements from 15 August 2016 observations of the variable S  star $\chi$ Cygni. In this epoch the telescope vibrations (and resulting large phase error) were mitigated by the implementation of the SCExAO low-order wavefront sensor (LOWFS), and accordingly the `double-hump' feature is no longer seen in the data. The star is resolved by the GLINT Nuller with an angular diameter of 20.5~mas measured from the self-calibrated null-depth, consistent with known values.
}
  \label{fig_results_chiCyg201608}
\end{figure} 
\begin{figure}
  \centering
  \includegraphics[width=0.5\textwidth]{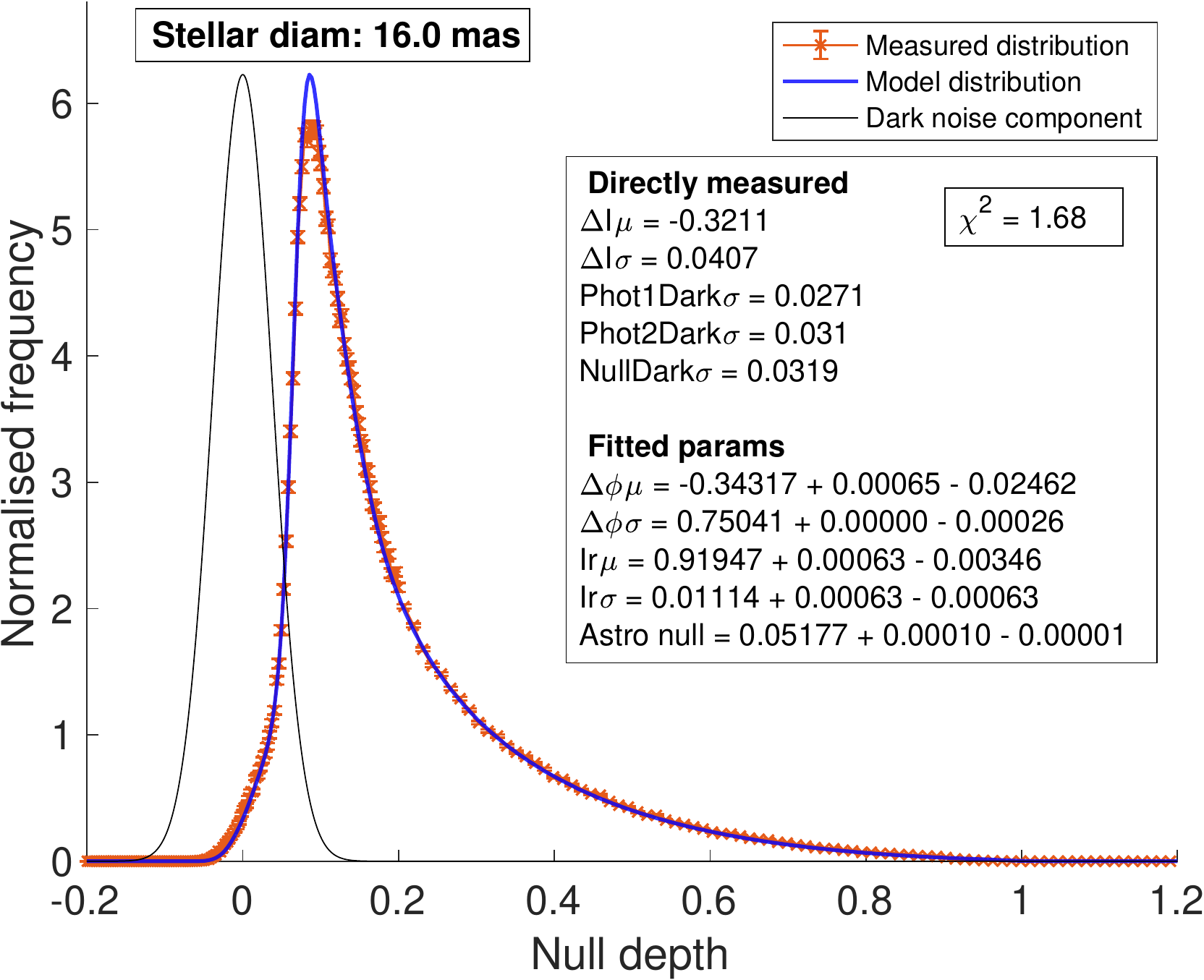}
  \caption{
  Histogram and fitted model PDFs of NSC null-depth measurements from 9 November 2016 observations of the K star $\alpha$ Tauri. Again, the telescope vibrations (and resulting large phase error) were mitigated by the implementation of the SCExAO low-order wavefront sensor (LOWFS). 
The star is barely resolved by the GLINT Nuller, with an angular diameter of 16.0~mas measured from the self-calibrated null-depth, slightly smaller than known values determined via interferometry. To improve clarity, data points at N>0.2 have been further binned for plotting.
}
  \label{fig_results_alfTau201611}
\end{figure}

\begin{figure}
  \centering
  \includegraphics[width=0.4\textwidth]{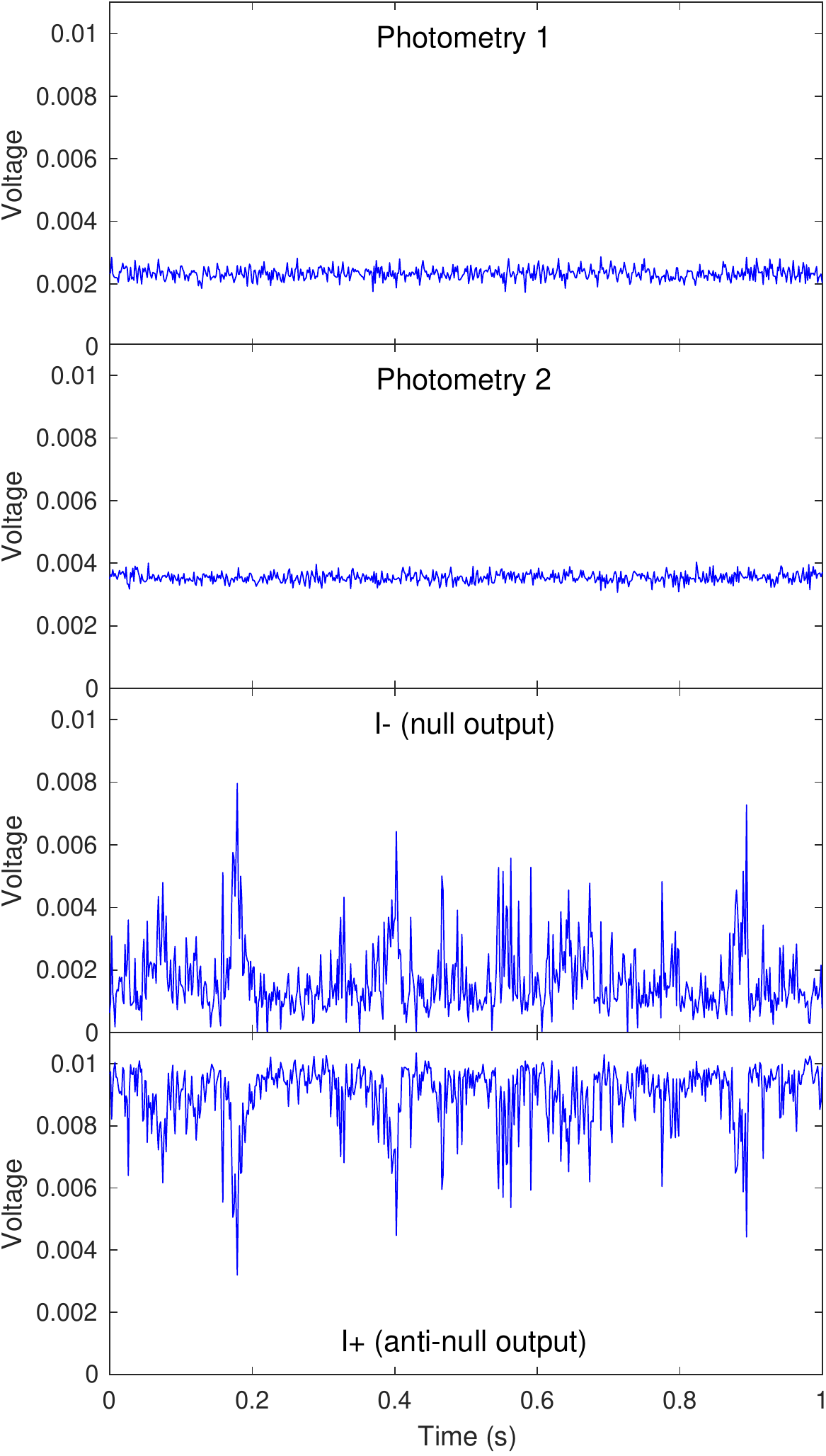}
  \caption{Raw time-series data from the 4 outputs of the chip for 1 second of on-sky observations of $\alpha$ Tauri (with processed data shown in Figure \ref{fig_results_alfTau201611}). The y-axis is in units of detector output voltage (proportional to channel intensity). This observation represented the best data in terms of RMS phase error, due to SCExAO's low order mode correction at this epoch.}
  \label{fig_timeseriesalftau}
\end{figure}

Observations of another partially-resolved star, the red giant $\alpha$ Herculis, were also performed during this same epoch, also subject to similar telescope vibrations. An astrophysical null depth of $0.1850\substack{+0.00019 \\ -0.00002}$ was measured, and the results are shown in Figure~\ref{fig_results_alfHer201603}. Despite the large phase errors the measured null depth provides a UD diameter of 31.2~mas, consistent with existing long-baseline interferometry measurements which give the K band UD diameter as between 31 and 33 mas \cite{Richichi2005, Duvert2016}.

For both of these vibration-affected observations, the fitted value of $\Delta \phi_\sigma$ is large: 2.3 radians for $\alpha$ Bootis and 1.8 radians for $\alpha$ Herculis. Furthermore, the magnitude of the relative intensity deviation $I_r$ (defined as the ratio of the `true' $I_+$ value to the estimate used) was large, with a mean $I_r$ of 0.85 for $\alpha$ Bootis and 0.70 for $\alpha$ Herculis. Having this deviation as a free parameter allows the model to successfully fit the large phase error and `double hump'. As will be seen, for low phase-error observations $I_{r \mu} \approx 1$. 

In a subsequent observing epoch (August 2016) the telescope vibrations were partly mitigated by the introduction of a low-order wavefront sensor (LOWFS) into SCExAO \citep{Singh2014}, which allowed low order spatial modes to be corrected at high gain without causing instability in higher-order modes. The variable S-type red giant star $\chi$ Cygni was observed at this time, with the resulting histogram shown in Figure \ref{fig_results_chiCyg201608}. It is clearly seen that the second `hump' is now absent, with a mean relative intensity deviation of almost unity ($I_{r \mu}$ = 0.98). It was found to have an astrophysical null depth of $0.0824\substack{+0.0004 \\ -0.0002}$ which corresponds to a UD stellar diameter of 20.5~mas. The closest-wavelength literature measurement is a UD diameter of 23.2~mas measured at K' band \citep{Mennesson2002b}. This is consistent with the diameter measured here, particularly given the strong dependence of diameter on wavelength for this star (e.g. it is 30.4~mas at L'), its highly extended atmosphere and known variability due to stellar pulsations. 

Additional observations in November 2016 of the K-type red giant star (and possible long period variable) $\alpha$ Tauri showed similarly small phase error, with the results shown in Figure \ref{fig_results_alfTau201611}. The measured astrophysical null depth of $0.05177\substack{+0.0001 \\ -0.00001}$ corresponds to a UD stellar diameter of 16.0~mas. This is smaller than published H band diameters of between 19.5 and 20.5~mas \citep{Richichi2005}. However it should be noted that this star is about 3 times smaller than the formal diffraction limit for this baseline (73~mas). Again there is no second `hump', with a mean relative intensity deviation approaching unity ($I_{r \mu}$ = 0.92). A time-series plot of 1 second of data for this observation is presented in Figure \ref{fig_timeseriesalftau}.

\begin{figure}
  \centering
  \includegraphics[width=0.5\textwidth]{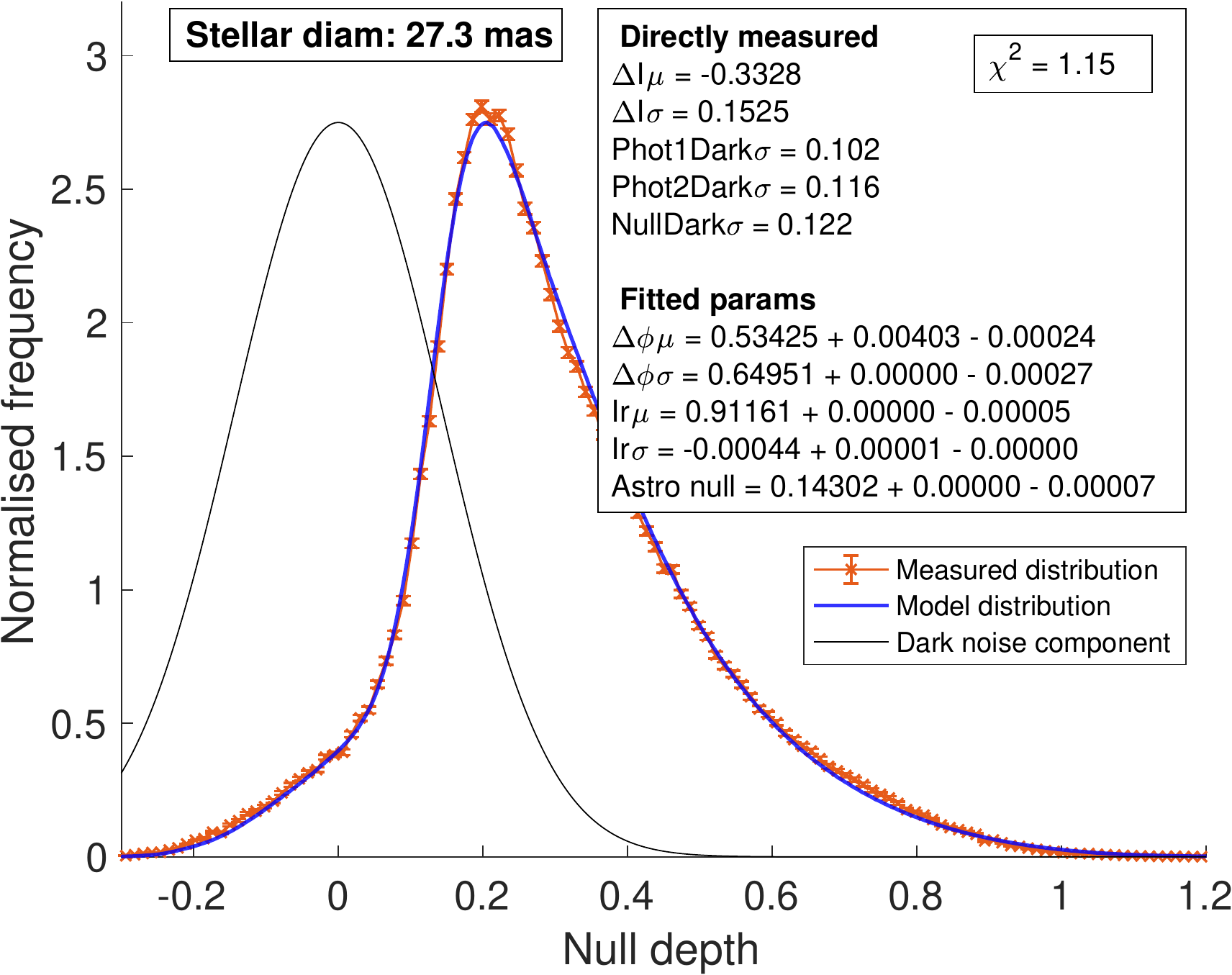}
  \caption{
  Histogram and fitted model PDFs of NSC null-depth measurements from 9 November 2016 observations of the Mira variable star o Ceti. Telescope vibrations are mitigated but extended wings appear in the histogram, with a pronounced `kink' around $N=0$. This is due to RF noise in the dark PDFs (shown in Figure \ref{fig_omiCetDarkHists}), which is successfully fit via the NSC method.  
The star has an angular diameter of 27.3~mas measured from the self-calibrated null-depth, consistent with known values.
}
  \label{fig_results_omiCet201611}
\end{figure}

\begin{figure}
  \centering
  \includegraphics[width=0.4\textwidth]{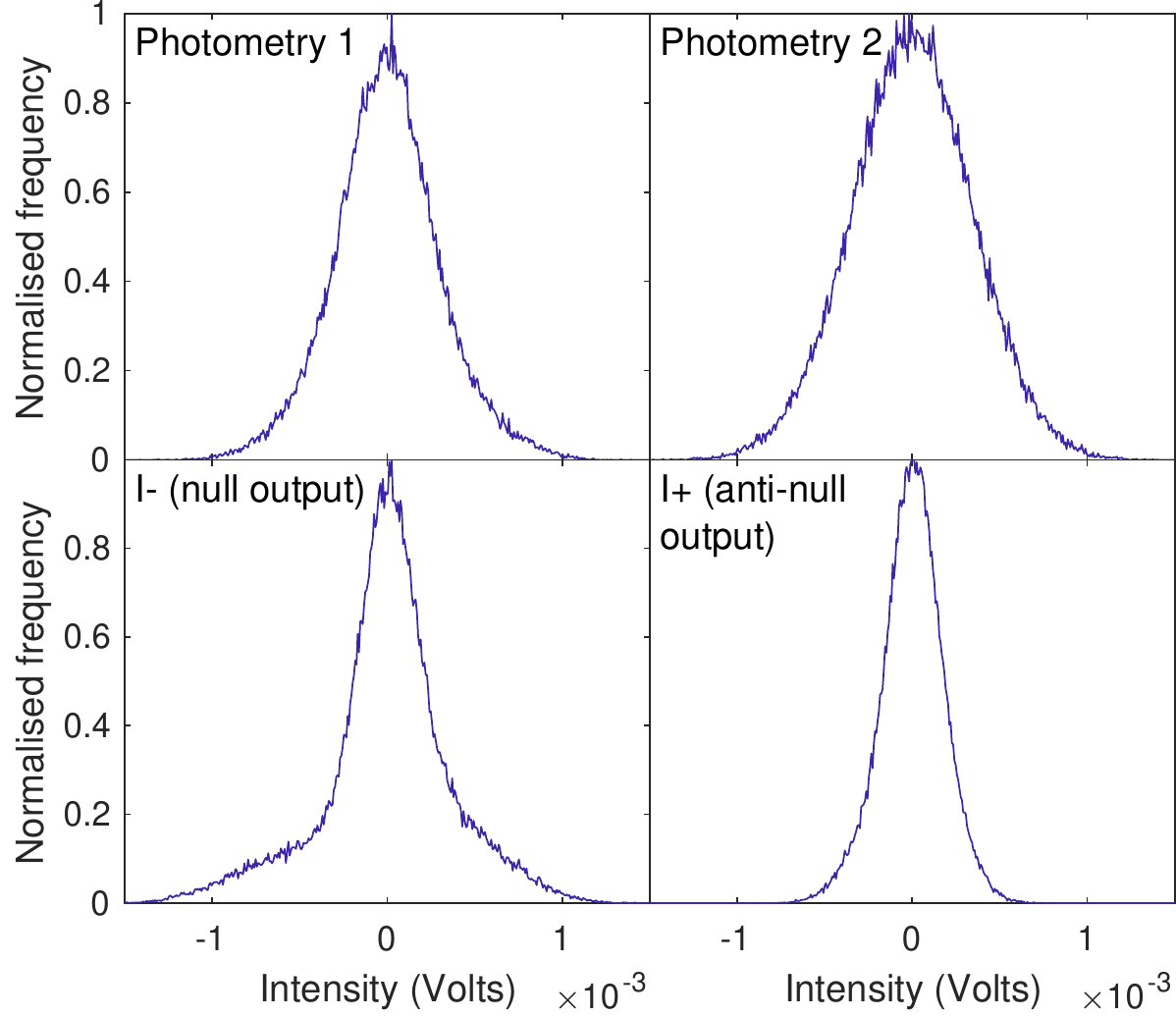}
  \caption{Histograms of the dark noise for the 4 detectors during the observations of o Ceti, the results of which are shown in Figure \ref{fig_results_omiCet201611}. The unusual non-Gaussian shape of the noise histogram of the null-channel is believed to be due to RF interference, and results in an extension to the wings of histogram in the final data. However since this PDF measurement is directly used by the model, the NSC method accounts for the problem still producing a good fit to the data.}
  \label{fig_omiCetDarkHists}
\end{figure}

Observations of the variable red giant star o Ceti (Mira) also exhibited no major vibration issues, and the results are shown in Figure~\ref{fig_results_omiCet201611}. The observed astrophysical null of $0.14302\substack{+0.00004 \\ -0.00007}$ corresponds to a UD diameter of 27.3~mas. This is consistent with the range of UD diameters measured for this star, with K band long-baseline interferometry measurements ranging between 27 and 33~mas reported \cite{Woodruff2008, Woodruff2009, Richichi2005, Duvert2016}. Additionally, the null-depth histogram in Figure \ref{fig_results_omiCet201611} is seen to have a curious shape, with the distribution having wide wings (with a `kink' around $N=0$) not seen in the other observations. The explanation for this can be seen by reference to the histograms of the dark-current for the 4 detectors, shown in Figure \ref{fig_omiCetDarkHists}. While 3 channels show the expected Gaussian distribution, the null channel features additional wings, which is partly due to mains-frequency noise pickup due to a poorly positioned cable. Nonetheless the model provides a good fit to the data, demonstrating the advantage of the NSC method (this PDF shape would not be reproduced by the model if a Gaussian distribution was assumed, as is done for the ASC approach).


\begin{figure}
  \centering
  \includegraphics[width=0.5\textwidth]{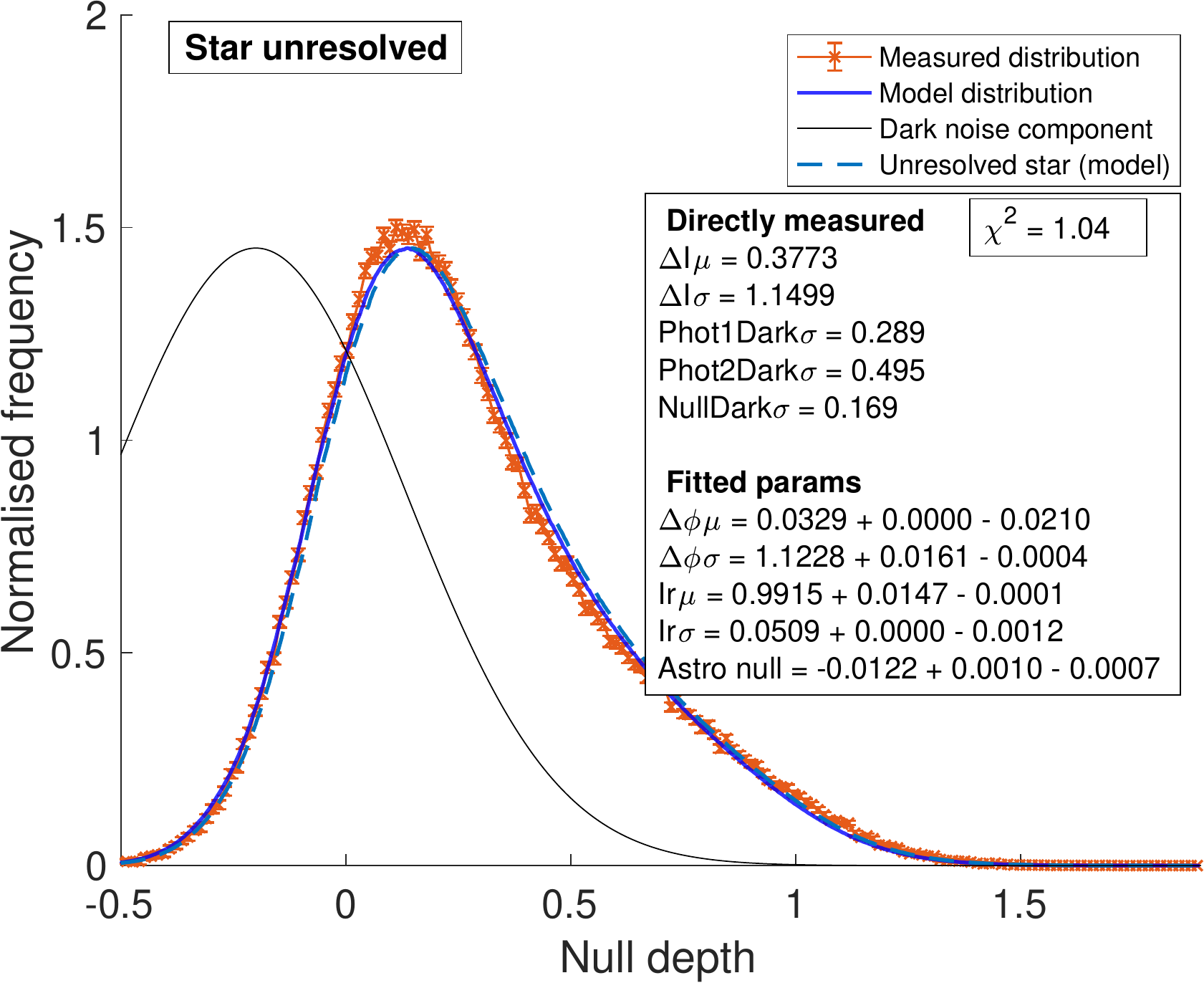}
  \caption{
 Histogram and fitted model PDFs of NSC null-depth measurements from 19 March 2016 observations of the unresolved star Vega. The contribution from dark noise is high, but the fitted model (blue line) is consistent with that of an unresolved star (broken light-blue line, overlapping). The fitted null depth should be zero, but is poorly constrained and slightly negative, likely due to the poor constraints on null depth due to the dominating dark noise.
}
  \label{fig_results_vega2016}
\end{figure}

Lastly, the results of an observation of the unresolved star Vega are shown in Figure \ref{fig_results_vega2016}. Although a bright star, this was the faintest in our sample and the low sensitivity of the detectors used meant that the contribution from dark noise was very large (as seen in the Figure). Since the star is unresolved the null depth should be zero, but a slightly negative null-depth (-0.012) is fitted, which is non-physical. However a large variability in the fitted null depth was observed between different model-fitting runs (with different initial positions), with a standard deviation between results of 0.01. It is likely the dark-noise dominated signal was unable to provide a more accurate constraint. Nonetheless within these limits it is consistent with an unresolved star, and is visually demonstrated by the over-plotting of an unresolved star in the Figure. A table summarising all the above observations is included as Table \ref{table_results}.

\begin{table}
\begin{tabular}{@{}lllll@{}}
\toprule
Obs. Date & Star              & $N_A$   & Fitted diam. & Literature diam.     \\
          &                   &         & (milliarcsec) & (milliarcsec)$^1$	  \\ \midrule
-         & Lab source        & 0.0083  & -            & -                    \\
19/03/16  & $\alpha$ Bootis   & 0.0705  & 18.9         & 19.1 - 20.4 		  \\
21/03/16  & $\alpha$ Herculis & 0.1850  & 31.2         & 31 - 33   			  \\
15/08/16  & $\chi$ Cygni      & 0.0824  & 20.5         & 23.2		          \\
9/10/16   & $\alpha$ Tauri$^2$& 0.0518  & 16.0         & 19.5 - 20.5          \\
9/10/16   & o Ceti            & 0.1430  & 27.3         & 27 - 33              \\
19/03/16  & Vega$^3$          & -0.0122 & \textless{}0 & 3                    \\ \bottomrule
\end{tabular}
\caption{Summary of stars observed and their fitted diameters. See text and individual figures for details. $^1$See text for references. $^2$For such under-resolved stars (here 3$\times$ smaller than the diffraction limit) the fitted diameter becomes inaccurate. $^3$The null depth here is poorly constrained, with different fitting runs giving results varying by $\pm 0.01$, likely because the dark noise is extremely high (see Figure \ref{fig_results_vega2016}). }
\label{table_results}
\end{table}

\section{Conclusion}
\label{sec_Conclusion}
Nulling interferometry is a promising method to directly image high contrast features at super-diffraction-limited angular resolutions, such as exoplanets and circumstellar disks. While the traditional approach is to use bulk-optics based interferometers, here we present an integrated-photonic approach. In this method all beam splitting and combination is performed within a single photonic chip, via single mode waveguides and evanescent directional couplers written within the substrate. The single-mode waveguides perform perfect spatial filtering, and the monolithic design promises extreme stability and compactness. Simultaneous photometry and anti-null outputs are easily implemented. Moreover, the integrated photonic approach allows complex interferometric devices -- such as those including a large number of baselines -- to be implemented easily.

The GLINT pathfinder instrument described here centres around a photonic nulling interferometry chip produced via the laser direct-write method. The design and implementation of the chip and surrounding instrumentation is described, and basic laboratory characterisation measurements are presented. The technique was validated via on sky deployment of the instrument at the 8~m Subaru Telescope using the SCExAO extreme adaptive optics system, and on-sky results are presented (analysed via the numerical self-calibration method). Despite high dark-current contributions (due to uncooled photo-detectors) and large dynamic phase errors (due to telescope vibration as well as AO residuals) the measured null depths successfully predict the stellar diameter of the observed stars, including cases where the stellar angular size is more than twice as small as the formal diffraction limit. Absolute null depths were limited by the chromatic nature of the device, with null depths of around 0.8\% found when using an unresolved laboratory source over a 50~nm bandwidth, consistent with expectations.

With the basic concept of the photonic nulling interferometer demonstrated, a number of key improvements will be made to the next iteration of the instrument. Firstly, more baselines will be utilised, allowing simultaneous measurements of the null depth of multiple baseline lengths and angles. This is important when measuring asymmetric sources, such as a star and faint companion (with coverage augmented via field rotation). This upgrade is straightforward due to the intrinsic scalability of the integrated photonic design. At first, 4 input waveguides will be used, producing 4 nulled baselines and 2 non-nulled baselines. Secondly, the sensitivity (dark-noise) issue will be addressed by replacing the uncooled photo-diodes with a sensitive, cooled imaging detector, such as a high performance InGaAs camera (e.g.  the CRed-2 \citep{Feautrier2018}) or e-APD array (e.g. a Saphira array \citep{Atkinson2018}).

Thirdly, the chromatic limitations will be addressed. Our approach is twofold. By utilising the aforementioned array detectors, the output fibres will be spectrally dispersed. By fitting a chromatic null-depth model to  measure null depth as a function of wavelength, precise astrophysical null depths can be measured. Additionally, the  directional couplers used in the chip can be made more achromatic using asymmetric designs (e.g. \cite{Chen2008}.

Finally, active wavefront control will be improved to produce deeper (and more stable) absolute nulls: a critical feature when suppression of stellar photon noise is required (such as in high-contrast companion detections). One approach is to improve the existing adaptive optics implementation, for example by using a low-order wavefront sensor close to the chip injection. This would reduce non-common path errors and could be tuned to aggressively correct the spatial modes corresponding to the baselines used. While this use of the extreme-AO system as an external fringe-tracker may be sufficient, it may also be desirable to include variable on-chip delays. These could be implemented with micro-heater technologies, wherein electrical heater elements are deposited on the surface of the chip and can actively modify the waveguide propagation constant. Alternatively the chip could be produced using materials such as Lithium Niobate, where on-chip electrodes vary the local refractive index of the waveguides via the electro-optic effect (e.g. \cite{Martin2014}). These methods allow phase control of individual waveguides with very high slew rates, and could be run in closed loop using the chip outputs as sensors.

While the pathfinder instrument operates in the near-IR (due largely to the maturity of photonic technologies in this region), it is scientifically optimal to conduct these observations in the mid-IR, at wavelengths of around $5 \mu$m. Here, the star/planet contrast is more favourable. To enable this, new technologies to produce mid-IR capable direct-write photonics are being developed. Due to the high opacity of normal silica glass at these wavelengths more exotic materials are required. Chalcogenide glasses, including GLS, have shown particular promise -- they are highly transparent at mid-IR wavelengths and have successfully been used to create waveguides and evanescent couplers using laser direct write \citep{Labadie2012, Arriola2014, Gross2015b}. Prototype nulling-interferometer beam-combiners have also been produced using direct-write with this glass \citep{Gretzinger2019}, as well as planar lithographic technology \citep{Kenchington2016}. Other materials such as various different types of fluoride and chalcogenide glasses and fluoride crystals have also been investigated \citep{Arriola2017}.

Together, we anticipate that these new technologies combined with the currently demonstrated pathfinder technology will lead to a new generation of integrated photonic nulling interferometers. As waveguide number and null depth performance increase, these instruments stand to be highly competitive in the direct imaging and spectroscopy of faint companions such as exoplanets, especially in ultra-high resolution applications (separations at or below $\lambda / D$) where other imaging methods (such as coronagraphs) are less effective.

\section*{Acknowledgements}
This work was supported by the Australian Research Council Discovery Project DP180103413. 
It was performed in part at the OptoFab node of the Australian National Fabrication Facility utilising Commonwealth as well as NSW state government funding. S. Gross acknowledges funding through a Macquarie University Research Fellowship (9201300682) and the Australian Research Council Discovery Program (DE160100714). N. Cvetojevic acknowledges funding from the European Research Council (ERC) under the European Union's Horizon 2020 research and innovation program (grant agreement CoG - 683029). The authors acknowledge support from the JSPS (Grant-in-Aid for Research \#$23340051$, \#$26220704$ \#$23103002$). This work was supported by the Astrobiology Center (ABC) of the National Institutes of Natural Sciences, Japan and the directors contingency fund at Subaru Telescope. This research was also supported by the Australian Research Council Centre of Excellence for Ultrahigh bandwidth Devices for Optical Systems (project number CE110001018). The authors wish to recognise and acknowledge the very significant cultural role and reverence that the summit of Maunakea has always had within the indigenous Hawaiian community. We are most fortunate to have the opportunity to conduct observations from this mountain.


\bibliographystyle{mnras}
\bibliography{bnorrisbib_GLINTPaper2018a} 




%
%


\bsp	
\label{lastpage}
\end{document}